\documentclass[prl,aps,showpacs,twocolumn,floatfix,10pt]{revtex4-1}
\usepackage{graphicx}
\usepackage{color}
\usepackage[normalem]{ulem}

\begin{document}

\author{V.\ M.\ Pudalov$^{a,c}$, M.\ E.\ Gershenson$^{b}$, H.~Kojima$^{b}$}
\affiliation{$^a$ P.~N.~Lebedev Physical Institute, 119991 Moscow, Russia}
\affiliation{$^b$ Department of Physics and Astronomy, Rutgers University,
New Jersey 08854, USA,}
\affiliation{$^c$ Moscow Institute  of Physics and Technology, Moscow 141700, Russia}
\date{\today}
\title{
 Probing Electron Interactions in a Two-Dimensional System by Quantum Magneto-Oscillations}

\begin{abstract}
We have experimentally studied the
renormalized effective mass $m^*$ and  Dingle temperature $T_D$
in  two spin subbands with essentially different
electron populations. Firstly, we found that the product $m^* T_D$ that  determines the damping of quantum oscillations, to the first approximation,  is  the same
in the majority and minority subbands even at a
spin polarization degree as high as 66\%. This result confirms the theoretical predictions  that the
 interaction takes place at high energies  $ \sim E_F$ rather than within a narrow strip of
energies $E_F\pm k_BT$.
Secondly, to the next approximation, we revealed a difference in the  damping factor  of the two spin subbands, which  causes skewness of the oscillation lineshape.
In the absence of the in-plane magnetic field  $B_\parallel$, the damping  factor $m^*T_D^*$
is systematically  {\em smaller in the spin-majority subband}. The difference,
quantified with the skew factor $\gamma= (T_{D\downarrow}-T_{D\uparrow})/2T_{D0}$  can be as large as 20\%.
The skew factor  tends to decrease as $B_\parallel$ or temperature grow, or $B_\perp$ decreases; for low electron densities and high in-plane fields
the  skew factor even changes sign.
Finally, we compared  the temperature and magnetic field dependences of the magnetooscillations amplitude
with predictions of the interaction correction theory, and  found, besides some qualitative similarities,  several quantitative and qualitative differences.
To explain qualitatively our results, we suggested an empirical  model that assumes the existence of easily magnetized  triplet  scatterers on the Si/SiO$_2$ interface.
\end{abstract}

\maketitle

\section{1.~Introduction}
\label{sec:intro}
The problem of interactions between electrons in a disordered two dimensional (2D)
electron system remains a field of active research
\cite{finkelstein, okamoto_1999, ZNA, vitkalov_2001, pud_gm_2002, savchenko_2002, kashuba_2002, pud_prl_2003, vitkalov_JPSJ_2002, shashkin_prl_2003, vitkalov_2003, savchenko_2003, pud_granada_2004, zhu, krav_review_RPP_2004, maslov_prl_2005, adamov_prb_2006, martin_prb_2003, spivak, khodel_2005, punnoose_science_2005, klimov_prb_2008, das_2005, clarke_nphys_2008}.
As the carrier density $n$ decreases, the electron-electron
interaction energy $E_{ee}$  exceeds by far
the kinetic Fermi energy   $ E_{ee}/E_F\sim r_s \gg 1$ \cite{note-interaction}.
The interactions reveal themselves in experiment via
renormalization of the observable quasiparticle parameters such as the
effective mass, spin susceptibility, $g$-factor, electron
compressibility, etc. \cite{manybody, pines, isihara_82, pud_gm_2002, clarke_nphys_2008}.

The interactions are usually treated within the framework of the Fermi-liquid (FL) theory, based on the concept of the low energy ( $\delta\varepsilon \ll E_F$ ) quasiparticles. The
FL theory states that the low-energy properties of an
interacting fermionic system are determined by the states in the
vicinity of the Fermi surface, and are similar to those of a weakly
interacting gas of quasi-particles with parameters  different from
 the bare band values. A great body  of data
supports this viewpoint even for $r_s$ values as high as $\sim 10$ for 2D electron systems.

A natural question arises whether the FL approach in 2D remains
valid as the interaction strength is further increased
\cite{anderson, khodel_2005, spivak}. The interaction is predicted to lead to various non-Fermi liquid ground states
\cite{khodel_2005, spivak,  tanatar-ceperley_1989, chui-tanatar_1995, chitra-giamarchi_2001}.
For example, it was suggested that in a  multicomponent electron liquid the interactions are mediated by exchange of
high-energy plasmons, irrelevant to spins, and thus should become similar to those in a bosonic liquid
\cite{kashuba_2002, punnoose_science_2005, maslov_prl_2005} as the number of components increases.

On the other hand, from intuitive expectations  based on the  RPA result \cite{overhauser_1971, maslov_prl_2005},   interactions within the minority subband should renormalize the effective mass stronger than  in the majority one.
In contrast to the latter expectations, earlier experiments \cite{note on mass, shashkin_prl_2003}  reported that the effective mass
extracted from the amplitude of  Shubnikov-de Haas (SdH) oscillations  in Si-MOSFETs
does not depend on the spin polarization to within 4-5\% accuracy.
These conclusions however were drawn from the temperature decay of the oscillation amplitude  which  was analyzed using the Lifshitz-Kosevich (LK) model for noninteracting electrons \cite{SdH} and by ignoring the difference between the subband parameters. Since then, the theory of quantum oscillations for an interacting 2D system has been developed \cite{martin_prb_2003, adamov_prb_2006};  it was also experimentally shown  that  the LK model is not fully adequate and its use leads to overestimated mass values \cite{klimov_prb_2008, pud_granada_2004}.

The Zeeman splitting of the spin subbands in the in-plane magnetic field can shed light on this puzzle.
 When the spin polarization $\zeta=(n_\uparrow-n_\downarrow)/n$ becomes of an order of 1, the spin subbands separation is of an order of $2E_F$. In this essentially high-energy problem,
which formally goes beyond the framework of the FL theory,
there is a  possibility to answer the question of how  electrons interact with each other in a
multicomponent system with different subband populations.
Particularly, whether interactions take place primarily within each spin-subband,
or in the entire electron system.

In the current work we address this question by performing precise measurements of the oscillations amplitude
and lineshape in independently controlled perpendicular and parallel magnetic fields.

The paper is organized as follows. Firstly, by using a model-independent general approach, we  compare the renormalized quasiparticle parameters
in  two unequally populated  spin subbands. By doing this, we verify the conclusion of the  earlier
experiments  \cite{pud_gm_2002, note on mass, shashkin_prl_2003} that
the electrons in the spin-minority and spin-majority subbands have almost the same ``damping factor'', i.e., the product of the effective mass
and  inverse quantum  time $m^*/\tau_q$, or, equivalently,
 $m^*T_D$ (where $T_D=\hbar/2\pi\tau_q$ is the  Dingle temperature). This equality holds with a reasonable accuracy, $\sim 15\%$,  even if the subband populations differ by more than 60\%.
 This experimental finding is  in line with the recent theory \cite{maslov_prl_2005}.

Secondly, we go beyond the qualitative  comparison,  and find that, to the next approximation,
the partial damping factor of the SdH oscillations,
 $m^*T_{D\uparrow}$,  in the majority (spin-up) subband
is systematically  {\em lower} (by $\sim 6 - 14\%$)
than that in the minority (spin-down) subband.
This is reminiscent of the predictions of the RPA model
\cite{maslov_prl_2005},
according to which for the majority subband  $m^*$  has to be smaller and $T_D$  to be lower due to better screening.

However, the overall picture is rather complicated:  the difference between the parameters
$(m^*  T_{D\uparrow\downarrow})$ in the two spin subbands diminishes  in the
limit of  weak perpendicular and strong parallel fields, or high temperatures. Moreover,
in strong parallel fields the relation between the $(m^*  T_{D\uparrow\downarrow})$ reverses,
and this observation seems to be at odds with the common-sense arguments based on the screening concept and  RPA.

Finally, we tested the theory of magneto-oscillations (MO) in the correlated 2D systems by analyzing  the temperature and field dependencies of the oscillation amplitude.
We reveal some similarities and some deviations from the theory.
Surprisingly, many of the  inconsistencies with theory vanish or weaken when a parallel field is applied in addition to the perpendicular field.

Our results outline the incompleteness of the existing theory of magneto-oscillations,
which, in our view,  should take into account on an equal footing not only the  temperature renormalization of the scattering rates and effective mass,
but also the  exchange-mediated interlevel interaction that modulates the energy splitting and screening,  intervalley scattering, valley splitting, and Zeeman effects in the in-plane field.

 In order to explain our results, we suggest an
  empirical model where  shallow  localized states at the Si/SiO$_2$ interface have a nonzero spin and can be easily spin polarized
in the external field; scattering of mobile electrons by the interface states modifies the magneto-oscillation line shape and amplitude.

\section{2. Experimental}
Our resistivity measurements were performed  with three high-mobility
Si-MOSFET samples \cite{sample} using a conventional four-terminal  ac technique.
Experiments have been done using two cross-field
superconducting coils \cite{crossed}.  This set-up allowed us to vary independently
the in-plane and perpendicular magnetic fields,
and  disentangle the electron parameters in the spin-up and spin-down subbands
by analyzing the
beating pattern (or lineshape) of the SdH oscillations.
The split-coil magnet generated the perpendicular magnetic field  that was used for observation of the SdH oscillations in both the spin-up and spin-down spin-subbands.
The in-plane magnetic field $B_\parallel$ was used to spin polarize the electron system; the difference between subband populations  is characterized by the spin polarization $\zeta = g^*\mu_B B_{\rm total}/2E_F$.
The Si-MOSFET samples are well suited for measurements in tilted fields due to the narrowness of the confining potential well, which minimizes the effect of the in-plane field on the orbital effects  \cite{stern_1968}.

Measurements were performed  in a dilution refrigerator over the temperature range  $0.1 -
1.2$\,K. This range corresponds mainly to the
ballistic regime of the electron-electron interactions, $2\pi kT\tau/[\hbar(1+F_0^a)] >1$ \cite{ZNA}.

\section{3. Theory of SdH oscillations in the interacting 2D Fermi liquid}
\label{sec:F-gas}
The  magneto-oscillations in the {\em noninteracting} Fermi gas are usually fitted by the  Lifshitz-Kosevich (LK) formula,
adapted for the 2D case and valid for
a small amplitude of oscillations $\delta\rho/\rho \ll 1$ \cite{SdH} :
\begin{eqnarray}
\frac{\delta\rho_{xx}}{\rho_0}&=& +2\frac{\delta D}{D}=\sum_i
A_i^{\rm LK}\cos\left[ \pi i\left(\frac{\hbar c \pi n }{e B_\perp}
-1\right)\right]
Z_i^s Z_i^v.
\nonumber\\
A_i^{\rm LK}&=&4\exp\left(-\frac{2\pi^2 i k_B
T_D}{\hbar \omega_c}\right) \frac{2\pi^2 i k_BT/\hbar \omega_c}{\sinh\left(\frac{2\pi^2 i
k_BT}{\hbar \omega_c}\right)}
\label{SdHbyLK}
\end{eqnarray}
Here  $\omega_c=eB_\perp/m^*c$ is the cyclotron frequency, $D$ is the 2D density of states,
$T_D=\hbar/2\pi k_B\tau_q$ is the Dingle temperature, $\tau_q$ is the
elastic all-angle (quantum) scattering time, and the Zeeman- and valley- splitting terms are:
\begin{equation}
Z_i^s=\cos \left[\pi i\frac{ \hbar \pi \zeta c n}{e B_{\perp}}\right], \quad
Z_i^v=\cos\left[ \pi i \frac{\Delta_V}{\hbar\omega_c}\right].
\label{zeeman_splitting}
\end{equation}

In the absence of the parallel field, the Zeeman term reduces to the field independent factor
$$
\cos\left(\pi i\frac{\Delta_Z}{\hbar\omega_c}\right)=\cos\left(\pi i\frac{g^*m^*}{2m_e}\right).
$$

In the case of {\em interacting} electrons, both the effective mass and the Dingle temperature are renormalized, which leads to an additional temperature  and magnetic field dependences
of the oscillations amplitude. The interaction quantum correction to the magnetooscillation amplitude  was considered in Refs.~\cite{martin_prb_2003, adamov_prb_2006} and  for the relevant case of Coulomb scattering reads as follows (Eq.~(86) in Ref.~\cite{adamov_prb_2006}):
\begin{eqnarray}
A1^{\rm int}_1 &=& A_1^{\rm LK} F_{\rm int} \nonumber\\
F_{\rm int} &=& \exp\left[\left\{1+15\frac{F_0^a}{1+F_0^a} \right\}\frac{\pi}{\omega_c\tau}\frac{k_BT}{E_F}\ln\left(\frac{\Delta}{T}\right)\right],
\label{eq:gornyi}
\end{eqnarray}
where $\Delta= q_{\rm TF}v_F = 4\pi e^2 D v_F/\overline{\kappa}$,  $D = 2m/\pi\hbar^2$, $\overline{\kappa}=7.7$ is the average dielectric constant \cite{ando_review}, the factor 15 is the effective number of triplet terms for a two-valley system in (100)Si-MOS, and the term in the figure brackets is the same as  in the quantum corrections to the conductivity   \cite{ZNA}.

\begin{figure}[ht]
\includegraphics[width=220pt]{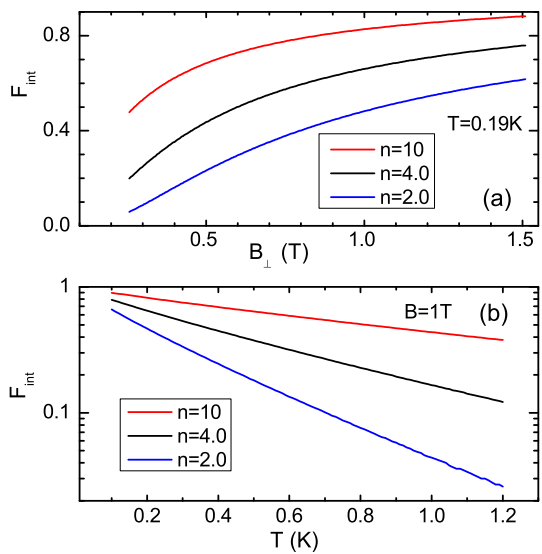}
\caption{(Color online) Calculated interaction corrections to the oscillation amplitude versus (a) $B_\perp$ field at $T=0.19$\,K,  and (b)  temperature for $B=1$\,T. Three representative density values are 2, 4, and 10  in units of $10^{11}$cm$^{-2}$.
 $F_0^a$ values equal to  -0.41, -0.334, and  -0.22, respectively \cite{klimov_prb_2008}.
 }
\label{fig:gornyi}
\end{figure}

Figure ~\ref{fig:gornyi} shows the magnetic field and temperature dependences  of $F_{\rm int}$ for three typical densities and relevant ranges of  temperatures and fields.

The extra  damping factor $F_{\rm int}$ is reduced to unity when the term in the figure brackets in Eq.~(\ref{eq:gornyi}) approaches zero; for a two-valley system with  15 triplets this should occur at $F_0^a=-0.0625$ (this value cannot be realized in conventional Si-MOSFETs). Therefore, for all accessible densities, the  interactions  are predicted to cause an additional decay of oscillations
with $1/B_\perp$  and with temperature.

This prediction can be explained as follows: in  the  LK model [Eq.~(\ref{SdHbyLK})],
 the amplitude damping factor contains the $T$-independent $T_D$:
\begin{equation}
-\ln\left[A_1^{\rm LK}(T,B_\perp) \right] \hbar \omega_c /2\pi^2 k_B \approx
\left(T+T_D \right).
\label{LK-damping}
\end{equation}
By contrast, in the interacting case [Eq.~(\ref{eq:gornyi})], the damping factor becomes temperature dependent:
\begin{equation}
-\ln\left[A_1^{\rm int}(T,B_\perp) \right]\frac{\hbar \omega_c}{2\pi^2 k_B}  \approx (T+T_D^*).
\label{gornyi-damping}
\end{equation}
Here, the effective $T_D^*$ includes corrections for both $m^*$ and $T_D$:
\begin{equation}
T_D^*= T_{D0}-\left(1+15\frac{F_0^a}{1+F_0^a}\right)
\frac{\pi}{\omega_c\tau} \frac{k_BT}{E_F}\ln\left(\frac{\Delta}{T} \right).
\label{reduced LK+gornyi}
\end{equation}

Equation (\ref{LK-damping}) is commonly used to find the effective mass from the slope of the $\ln A_1 (T)$  dependence. The interaction correction
to the damping, the second term in the latter equation,  is usually
negative and  enhances oscillation damping with temperature.
Its temperature dependence is nearly linear in $T$ [see Fig.~\ref{fig:gornyi}(b) and also Fig.~\ref{fig-n10_B0}(b)].
Therefore, the use of Eq.~(\ref{LK-damping})
instead of Eq.~(\ref{reduced LK+gornyi}) typically leads to an overestimated effective mass \cite{pud_granada_2004}. The interaction effects on the $T$- and $B_\perp$-dependencies of the MO amplitude  are very strong for two-valley systems, as  Figs.~\ref{fig:gornyi}(a)  and (b) show.

The renormalized mass $m^*$ and $g^*$-factor were determined
earlier in a number of experiments \cite{pud_gm_2002, pud_granada_2004, klimov_prb_2008};
these parameters  grow
monotonically as the density decreases (and $r_s$ increases) \cite{note-renormalized}.
This trend is in line with the expectation of the FL theory.
The Dingle temperature $T_D=\hbar/2\pi \tau_q$
is sample specific; for Si-MOSFETs it is (roughly) close to that determined by the transport scattering rate  $1/\tau_q\sim 1/\tau$.
Below we refer to this well established (though not fully  quantitatively explained) behavior  of the ensemble
averaged $T_D$, $m^*$ and $g^*$ for an unpolarized or partially polarized system
 as  ``conventional''.

 In what follows, we will use these known parameters and  Eqs.~ (\ref{reduced LK+gornyi}) and (\ref{LK-damping}) to find the effective $T_D^*(T,B_\perp, B_\parallel)$ from the measured oscillation amplitude
 at various $T$,  $B_\perp$  and $B_\parallel$. The values of  $T_{D\downarrow\uparrow}^*$ for individual subbands will be compared with each other, whereas the averaged value $T_D^*$ will be compared with  theoretical predictions.

\section{4. Oscillations  in the absence of in-plane field}

\subsection{4.1. High density, weak interaction, and simple oscillation spectrum}
\label{sec:high-density}
 We begin our analysis with oscillations at $B_\parallel =0$ and high electron densities $n\approx 10^{12}$/cm$^2$ ($r_s\approx 2.6$). For such a high density the effective mass is  well known \cite{fang_prb_1968, okamoto_1999, pud_gm_2002, note-renormalized}. On the other hand, this density is lower than the value $4\times 10^{12}$/cm$^2$, at which the second quantization subband gets populated \cite{ando_review}. The intersubband scattering therefore can be neglected and the electron system is truly two-dimensional. Additionally, since the Zeeman splitting at such high densities is much smaller than the cyclotron one,  the spectrum of  oscillations is simple with the first harmonic dominating.

\begin{figure}[ht]
\includegraphics[width=240pt]{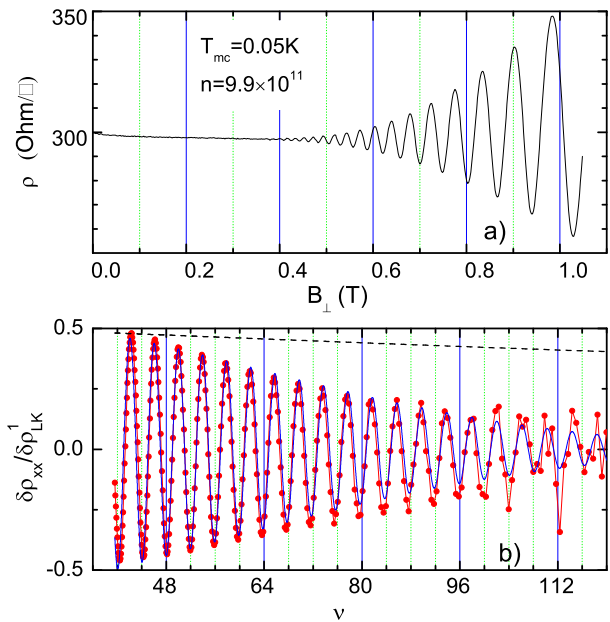}
\caption{(Color online) (a) The SdH oscillations $\delta\rho_{xx}$  vs  $B_\perp$ in the absence of the parallel field.
 (b)  The same oscillations  normalized to the calculated amplitude of the first harmonic, Eq.~(\ref{SdHbyLK}), plotted  vs the filling factor $\nu\propto 1/B_\perp$.   Dots with connecting lines are the data, the continuous curve -- calculations with an empirical field-dependent $T_D$ (see below) with two  parameters
 $T_{D0}=0.737$K,
 and $d_1 =0.1$T. The dashed line shows how the amplitude should vary according to Eq.~(\ref{eq:gornyi}). The canonical values for $g=2.57$ and $m^*=0.225 m_e$ have been used \cite{note-renormalized}.
 The carrier density is in units of cm$^{-2}$, $r_s=2.61$.
 }
\label{fig-n10_B0}
\end{figure}

An example of the  MO data $\rho_{xx}(B_\perp)$ is shown in Fig.~\ref{fig-n10_B0}(a). In order to  extract the oscillatory component $\delta\rho_{xx}$ from the raw data, we first cut off the low-field (weak localization, non-oscillatory) portion of the dependence  from 0 to about 0.15\,T, and then subtracted the  monotonic background magnetoresistance $\delta\rho_{xx}= \rho_0[1-\alpha(T,n)\times(\omega_c\tau)^2/(\pi \sigma_D e^2/h)]$ \cite{adamov_prb_2006} where we used $\alpha(T,n)$ as a fitting parameter [$\alpha(T)$ dependence is weak in the explored range of temperatures 0.1 -- 1K].

Further, in order to assign equal weights to all oscillations and for the simplicity of analysis, we normalize throughout the paper the extracted MO to the calculated amplitude of the first harmonic of Eq.~(\ref{SdHbyLK}),  $\delta\rho_{\rm norm} = \delta\rho/\rho_1^{\rm LK}$, where we set the Zeeman and valley factors in Eq.~(\ref{zeeman_splitting}) to unity. An example of oscillations with normalized amplitude is shown in Fig.~\ref{fig-n10_B0}b where they are plotted  as a function of the filling factor  $\nu=n hc/eB_{\perp}$ \cite{phase}. The reduced amplitude of the normalized oscillations here is simply the consequence of  smallness of the Zeeman factor, $Z_1^s =0.618$ for $g^* m^*/2m_e =0.288$.

\subsection{4.2. Experimental approach}
\label{sec:approach}
The period of oscillations is not renormalized in the presence of electron-electron interactions since it depends solely on the ratio of the electron density to the magnetic flux density -- neither of them is renormalized by interactions \cite{gorkov}.
The quasiparticle parameters, such as $g$ factor and $m^*$ are significantly renormalized even at such a high density, $r_s=2.61$.
The amplitude damping of $\exp$- and $\sinh$- functions is therefore affected by interactions and will be analyzed in the subsequent sections. Since one of our goals is to test the theory of magneto-oscillations,
we cannot apply the commonly used procedure  of disentangling $m^*$ and $T_D^*$ by using the so-called ``Dingle plot'', i.e. by plotting  $\ln(\delta\rho/\rho)$ versus $1/T$. Indeed, this procedure relies on Eq.~(\ref{SdHbyLK})[and Eq.~(\ref{LK-damping})], which  is  {\em a priori}  inapplicable.

In what follows we use a different approach: we compare the measured oscillation pattern with Eqs.~(\ref{SdHbyLK}), (\ref{zeeman_splitting}), and (\ref{eq:gornyi}) for a given temperature (and  $B_\parallel$   in the following sections), using earlier measured ``canonical'' values for the renormalized effective mass and $g$-factor \cite{note-renormalized}, and treat the Dingle temperature $T_D$ as an adjustable parameter.

\subsection{4.3 Over- and under- decay of the oscillations amplitude with inverse $B_\perp$ field}
When oscillations are normalized to the  amplitude of the first harmonic  calculated  from Eq.~(\ref{SdHbyLK}), their amplitude  is expected to be field independent for the non-interacting case. However, as Fig.~\ref{fig-n10_B0}(b) shows, this is not the case:  the MO amplitude at the lowest temperatures decays with inverse field faster than the LK formula predicts. Specifically, the MO amplitude in Fig.~\ref{fig-n10_B0}(a) over the range  $\nu=40-120$ (i.e., $B=1$ to $0.34$T) is expected to vary by a factor of 100, whereas in fact it drops by a factor of $\sim 400$. We note that the same extra damping may be found in earlier papers (see, e.g., Figs.~1(a) and 1(b) in Ref.~\cite{klimov_prb_2008}) though have never been discussed.  This extra damping is much stronger than the calculated from Eq.~(\ref{eq:gornyi}) interaction damping factor $F_{\rm int}$,  which varies only by 16\%  (from 0.917 to 0.77) over the field range.

To overcome the limitations of
Eqs.~(\ref{SdHbyLK}) and ({\ref{eq:gornyi}), we proceed
in the following  way:
we introduce an empiric field dependence of the Dingle temperature using an additional  adjustable parameter:
\begin{equation}
T_D=T_{D0}(1 + d_1/B_\perp).
\label{eq:empiric_TD(B)}
\end{equation}
In particular, to fit the oscillations
in Fig.~\ref{fig-n10_B0}(b) we have used  $d_1 =0.1$\,T; as a result $T_D$  varies by 15\% in the shown field range
and reasonably fits the data.
Hypothetically, the actual electron temperature (which is not measured independently) could be  higher than the known mixing chamber temperature. Under this assumption, the data  in Fig.~\ref{fig-n10_B0}(b)
may be  also fitted if the electron temperature is assumed to be 0.3K  (whereas the mixing chamber temperature is 0.05K).
The latter explanation, however,  is not supported by the data, because this ``extra" decay almost vanishes when the in-plane field is applied in addition to $B_\perp$   (see below).

As  temperature increases, the ``over-decay'' weakens and eventually transforms into an ``under-decay'', i.e., the oscillation amplitude starts decaying with the inverse field  {\em weaker} than the LK formula predicts. This deficiency is not as strong as the excessive decay discussed above: for example, in Fig.~\ref{fig-n10_B0_T0.5}, the raw oscillation amplitude $\delta\rho_{xx}$ at $T=0.5$K drops by a factor of 300 within the shown range $\nu=32 - 128$ (i.e. $B_\perp=1.2 -  0.3$\,T).
The normalized oscillations grow by  a factor of 2;
this growth is modeled in Fig.~\ref{fig-n10_B0_T0.5} by the field dependent $T_D=0.6(1+ d_1/B)$K (now with a negative  $d_1=-0.04$\,T).
Similarly,
the oscillations for the same density at $T=1$\,K  have been   fitted  with negative $d_1= -0.09$\,T. The  ``under-decay''  is qualitatively reproducible in different cooldowns and, in contrast to the low-temperature data,  cannot be explained  by electron overheating. Obviously, the ``under-decay'' cannot  be explained by interaction correction Eq.~(\ref{gornyi-damping}), which  may  produce only an extra decay with an increase of $1/B_\perp$;
for this particular temperature the predicted correction varies by a factor of 2 in the range of fields shown in Fig.~\ref{fig-n10_B0_T0.5}.

The over- and underdecay also cannot be simply caused by a mistake in the calculated amplitude of oscillations;
neither can it be attributed to the carrier  density inhomogeneity over the sample area, nor to any instrumentation  error. The amplitude variations are  strong (up to a factor of  4);   therefore they are most likely related to the exponential damping factors in Eq.~(\ref{SdHbyLK}) rather than to the Zeeman and valley cosine terms in Eq.~(\ref{zeeman_splitting}).

\begin{figure}[ht]
\includegraphics[width=230pt]{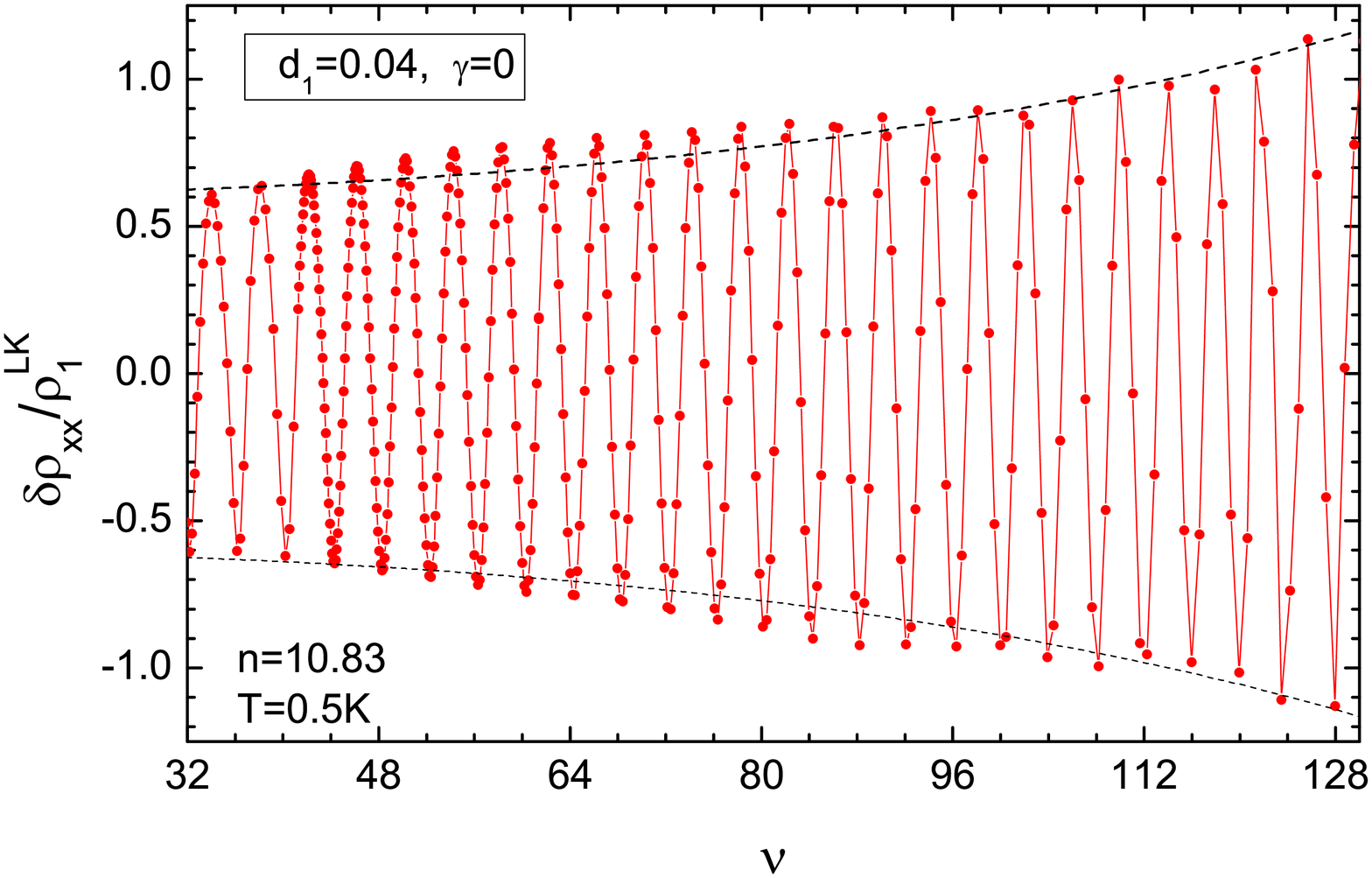}
\caption{(Color online) The SdH oscillations $\delta\rho_{xx}/\rho_1^{LK}$  normalized to the amplitude of the first harmonic calculated with a field independent $T_D=0.61$K,
 vs  filling factor  $\nu\propto 1/B_\perp$  in the absence of the parallel field.
 Dots with connecting lines are the data, the dashed curves show the calculated oscillations envelope with empirical field-dependent $T_D=0.6(1-0.04/B_\perp)$K. The canonical values for $g=2.57$ and $m^*=0.225$ have been used \cite{note-renormalized}.
 The carrier density is $10.83\times 10^{11}$cm$^{-2}$ and $T=0.5$\,K.
 }
\label{fig-n10_B0_T0.5}
\end{figure}

For completeness, we note that  there is a third option to explain the (not-too-strong) over- and under- decay, by assigning different $T_D$ values to the spin-up and spin-down subbands.
For example, the data in Fig.~\ref{fig-n10_B0_T0.5}  may be  well fitted  with $d_1=0$ and  the use of $T_{D\uparrow,\downarrow}=0.62(1 \pm \gamma)$K with $\gamma=0.18$.

For small Zeeman splitting $g\mu_B B \ll \hbar\omega_c$, the oscillations are featureless and we cannot determine unambiguously which of the abovementioned mechanisms is relevant. However, when the Zeeman splitting becomes comparable to the  cyclotron one (which occurs either at low densities, or in the presence of $B_\parallel$), as will be shown below,
the assumptions of the relevance of  $d_1$ and $\gamma$ gain
solid experimental ground.
Moreover, as will be  shown below, the over-decay   disappears when a rather weak $B_\parallel$ field is applied  in addition to the $B_\perp$ field.

\subsection{4.4. On the apparent smallness of the interlevel exchange interaction}
In Fig.~\ref{fig-n10_B0}, the peak-to-peak variations  $\delta\rho/\rho$ reach $\approx 28\%$ at 1.3\,T; correspondingly, the  oscillations of the density of states $\delta g/g =1/2(\delta\rho/\rho)$ reach $\approx 14$\%.  The latter should cause oscillatory level broadening \cite{pud_jetp_1985, DOS_oscillations}. In order to test its importance,  following Ref.~\cite{isihara_82}, we modeled level broadening  with
\begin{equation}
T_D(\nu)=T_{D0}(1-\delta g/g).
\label{TD_ocs}
\end{equation}
It appears, however,  that the introduction of the  oscillatory $T_D(\nu)$ dependence does not eliminate the extra decay of the  amplitude; we conclude that this factor is irrelevant.

The  underdecay may also be caused by enhancement of the oscillations due to the interlevel exchange interaction \cite{amplitude_enhancement, macdonald_1986}. This effect is caused by the exchange energy contribution, which lowers the chemical potential in the vicinity of integer fillings  and makes the thermodynamic density of states  negative \cite{efros_ssc_1988,  DOS_oscillations}:
\begin{equation}
\frac{\partial n}{\partial \mu} = - \alpha \frac{\kappa}{e^2 l_H} \left\{
\begin{array}{cl}
\widetilde{\nu}^{1/2} \quad  {\rm for } \quad \widetilde{\nu}\leq 1/2;\nonumber\\
(1-\widetilde{\nu})^{1/2} \quad  {\rm for } \quad \widetilde{\nu}> 1/2
\end{array}
\right.
\end{equation}
Here $l_H$ is the magnetic length,  $\widetilde{\nu}$ is the fractional part of the filling factor, and $\alpha $ (which is of an order of 1 for classical Coulomb interaction) was experimentally found to be about 0.04 -- 0.06 for Si-MOSFETs\cite{amplitude_enhancement}.
The estimated exchange contribution to $\partial n/\partial \mu$ is 10 times smaller than the single particle density of states and therefore is expected to cause at most a 20\% correction  to the amplitude of $\rho_{xx}$ oscillations. We conclude therefore that the exchange enhancement of cyclotron splitting cannot be (solely) responsible for the factor of 4 extra decay of oscillations.

\subsection{4.5. Proximity to the ``spin-zero'' regime}
In the absence of $B_\parallel$ field, the spin subbands are created  by the $B_\perp$ field, and according to the LK formula Eq.~(\ref{SdHbyLK}), the $g^*$ factor affects only their magnitude provided the Zeeman splitting is $\lesssim T_D$. One might think therefore that the oscillations are featureless,  insensitive to the quasiparticle parameters and   renormalization in individual spin subbands. However, as we show below, even in this case, the oscillation line shape provides information on the quasiparticle spectrum. This opportunity appears at lower electron densities, where the renormalized spin susceptibility $g^* m^*/2m_e$ approaches $1/2$ \cite{pud_gm_2002, okamoto_1999}. This situation corresponds to the  proximity of the spin splitting to half the cyclotron splitting. Under such conditions (called the ``spin-zero'' case),  $Z_1^s$  tends to vanish, which causes decrease of the main harmonic of the oscillations.

Figures \ref{fig-n4.23_B0} and \ref{fig-n4.14_B0}  show two examples of oscillations and their modeling with the LK formula Eq.~(\ref{SdHbyLK}).
The oscillations significantly deviate  from the harmonic function.
The corresponding data have been measured at two different cooldowns and the similarity of  the features discussed below
confirms that they are not caused by inhomogeneity of the electron density  in the 2D system.
Similar to the data  in Fig.~\ref{fig-n10_B0}, for such low fields, the main period of oscillations $\Delta\nu = 4$ reflects the four-fold spin- and valley degeneracy of the 2D electron system in (100)-Si.
The amplitude of the oscillations is  less than unity, which again suggests that  the Zeeman splitting is close to the half the cyclotron splitting in Eq.~(\ref{zeeman_splitting}),  i.e. $(g^*m^*/2m_e)=0.44$ is close to $1/2$ for which $Z_1^s=0$.   This agrees with the directly measured  renormalized spin susceptibility  $(g^*m^*/2m_e) =  1.9m_b=0.39$ for the given density value $r_s\approx 4$,  Ref. \cite{pud_gm_2002}.

Similar to the discussed case of higher densities,  the amplitude of the {\em normalized}  oscillations  in Figs.~\ref{fig-n4.23_B0} and  \ref{fig-n4.14_B0}   is decaying monotonically versus the inverse  field (by a factor of 4 in the shown field range), rather than being  field independent according to  the LK model Eq.~(\ref{SdHbyLK}). This effect is ten times stronger than that expected due to the interaction correction (Fig.~\ref{fig:gornyi}).
This  justifies our  empirical assumption  of the field-dependent level broadening $T_D(B_\perp$) [see Eq.~(\ref{eq:empiric_TD(B)}) and Fig.~\ref{fig-n10_B0}].

\subsection{4.6. Asymmetry of  two spin subbands}
In the magneto-oscillation dependencies  shown in Figs. \ref{fig-n4.23_B0} and \ref{fig-n4.14_B0},  there is a
``dip''   due to the emerging Zeeman splitting at $\nu\approx 14$. When the splitting is not fully resolved yet,
the dip position depends primarily on the $g^*$ factor  and points to  its enhanced value,
$g^*=3.4$, as compared to the ``canonical''  value 3.15 \cite{note-renormalized} measured in Refs.~\cite{pud_gm_2002, klimov_prb_2008} in the presence of a non-zero $B_\parallel$ field.

The asymmetry of the dip
shoulders seen in Figs.~\ref{fig-n4.23_B0}(b), \ref{fig-n4.14_B0}(b),  and \ref{fig-n6.18_B0} suggests some nonequivalence of the two spin subbands. To explain the asymmetry,  the oscillation magnitudes produced by the individual subbands must be different. The amplitude of oscillations is controlled by the product $m^*T_D$ in Eq.~(\ref{SdHbyLK}); its variations for different subbands may be due to the difference in either  $m^*$ or $T_D$ values.
The  estimate for the  polarization dependence of the effective mass
calculated  in Ref. \cite{maslov_prl_2005} in the  large -$N$ approximation [the degeneracy $N=4$ in (100) Si-MOS]:
\begin{equation}
m^*_{\downarrow\uparrow}/m^* =
\frac{1+\zeta^2}{12N}\log\left[\frac{r_sN^{3/2}}{1+\zeta^2}  \right]
\label{RPA estimate_of skewness}
\end{equation}
is negligibly small,  e.g., the  mass changes by less than 1\% when $\zeta$ varies from 0 to 1 for  $n=4\times 10^{11}$cm$^{-2}$ ($r_s=4$).
It is stressed in Ref.~\cite{maslov_prl_2005} that there is no spin splitting of the effective mass within the large-$N$ approximation.
 We therefore assume that $m^*$ is the same for both subbands and attribute the difference in amplitude to the difference in Dingle temperatures. Correspondingly,
 we model the asymmetry using
two different $T_D$ values, for the majority and minority subband:
\begin{eqnarray}
T_{D\uparrow} &=&(1-\gamma)  \quad {\rm {majority \quad subband}}\nonumber\\
T_{D\downarrow} &=& (1+\gamma) \quad {\rm {minority \quad subband}}.
\label{skew_factor}
\end{eqnarray}

The dip asymmetry  allows us to determine the skew factor $\gamma$ at those $B_\perp$ fields that correspond to the spin gaps (or  beating nodes for nonzero $B_\parallel$, which are discussed below). The resulting $\gamma$  appears to be  unexpectedly large; it is also cool-down dependent, in contrast to the $g^*$-factor:  $\gamma  \approx 0.18$  for Fig.~\ref{fig-n4.23_B0}, and  $\gamma \approx 0.12$  for Fig.~\ref{fig-n4.14_B0} (the two measurements have been performed at about the same density and with the same sample, but in two different cool-downs).
The different depth of the
dip at $\nu\approx 14$ in Figs.~\ref{fig-n4.23_B0} and \ref{fig-n4.14_B0} indicates irreproducibility of the $\gamma$ value in different cooldowns.

\begin{figure}[htp]
\includegraphics[width=230pt]{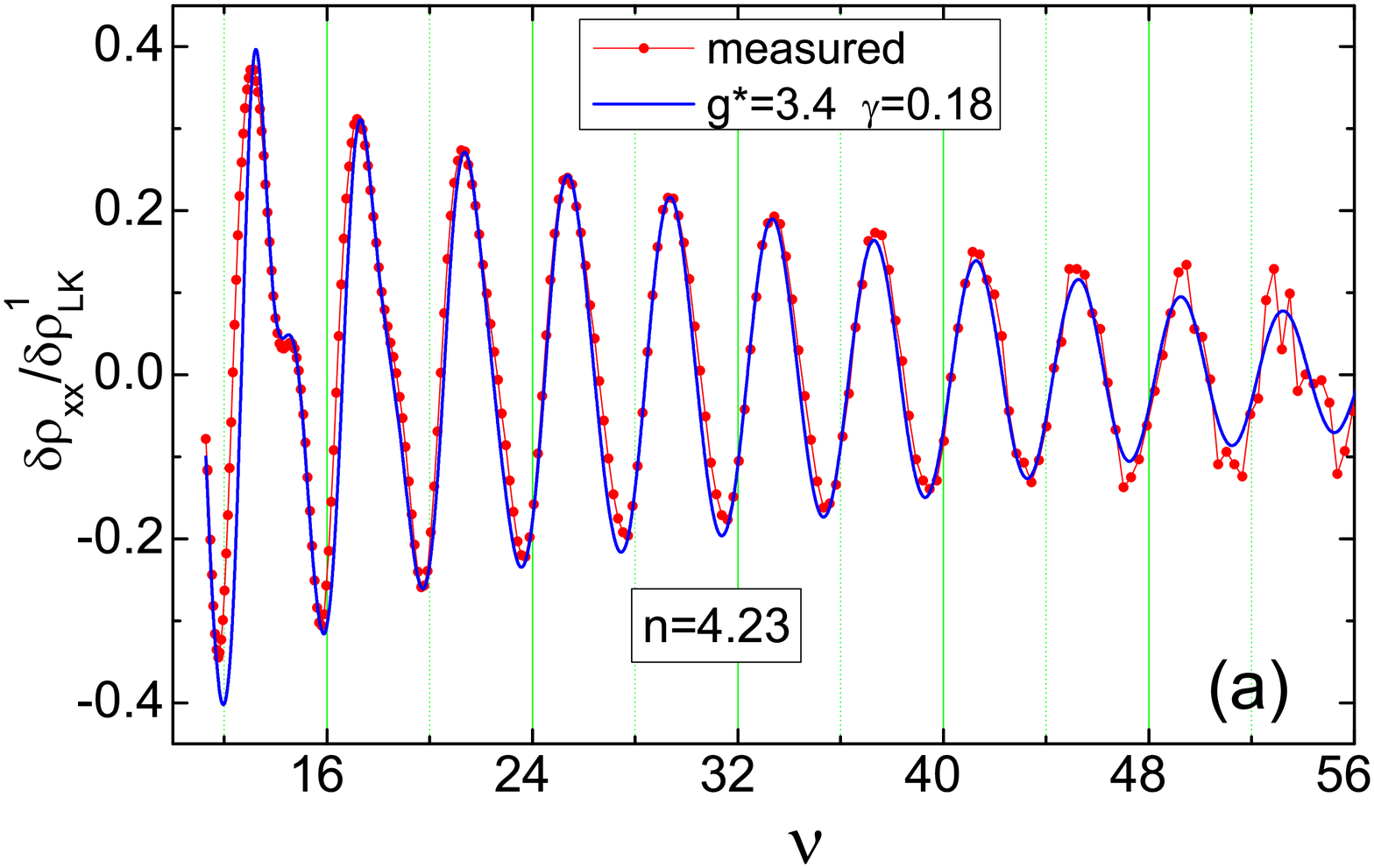}
\includegraphics[width=230pt]{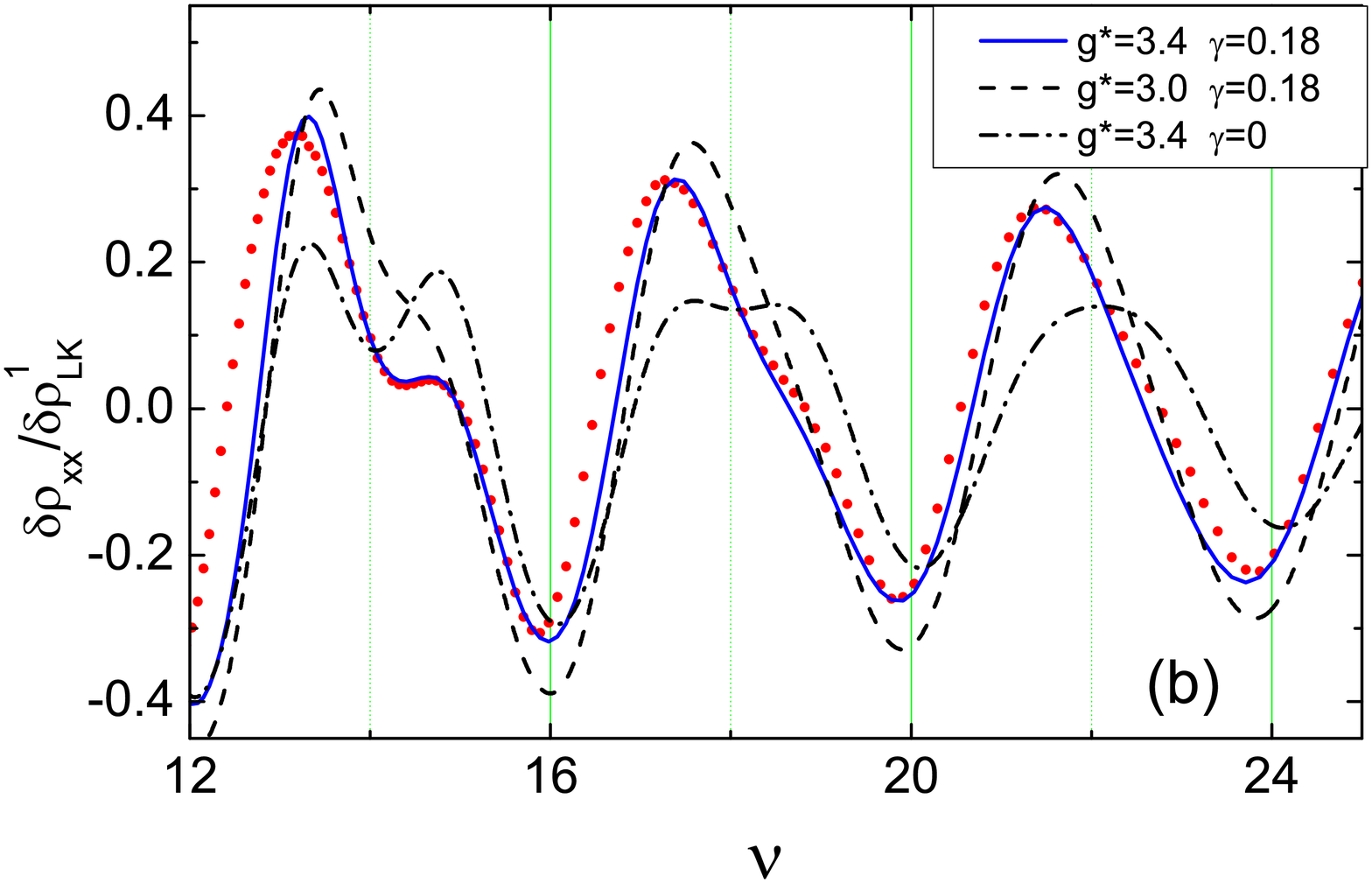}
\caption{(Color online) Normalized SdH oscillations $\delta\rho_{xx}/\delta\rho_1^{LK}$ in the absence of the parallel field vs  filling factor $\nu\propto 1/B_\perp$. $T=0.4$K,  carrier density
 $4.23\times 10^{11}$cm$^{-2}$  ($r_s\approx 4.0$)]. (a) over a wide range of fields $B_\perp= 0.3 - 1.5$\,T, and (b) oscillation fits corresponding to the sets of parameters shown in the figure.
 }
\label{fig-n4.23_B0}
\end{figure}

\begin{figure}[htp]
\includegraphics[width=230pt]{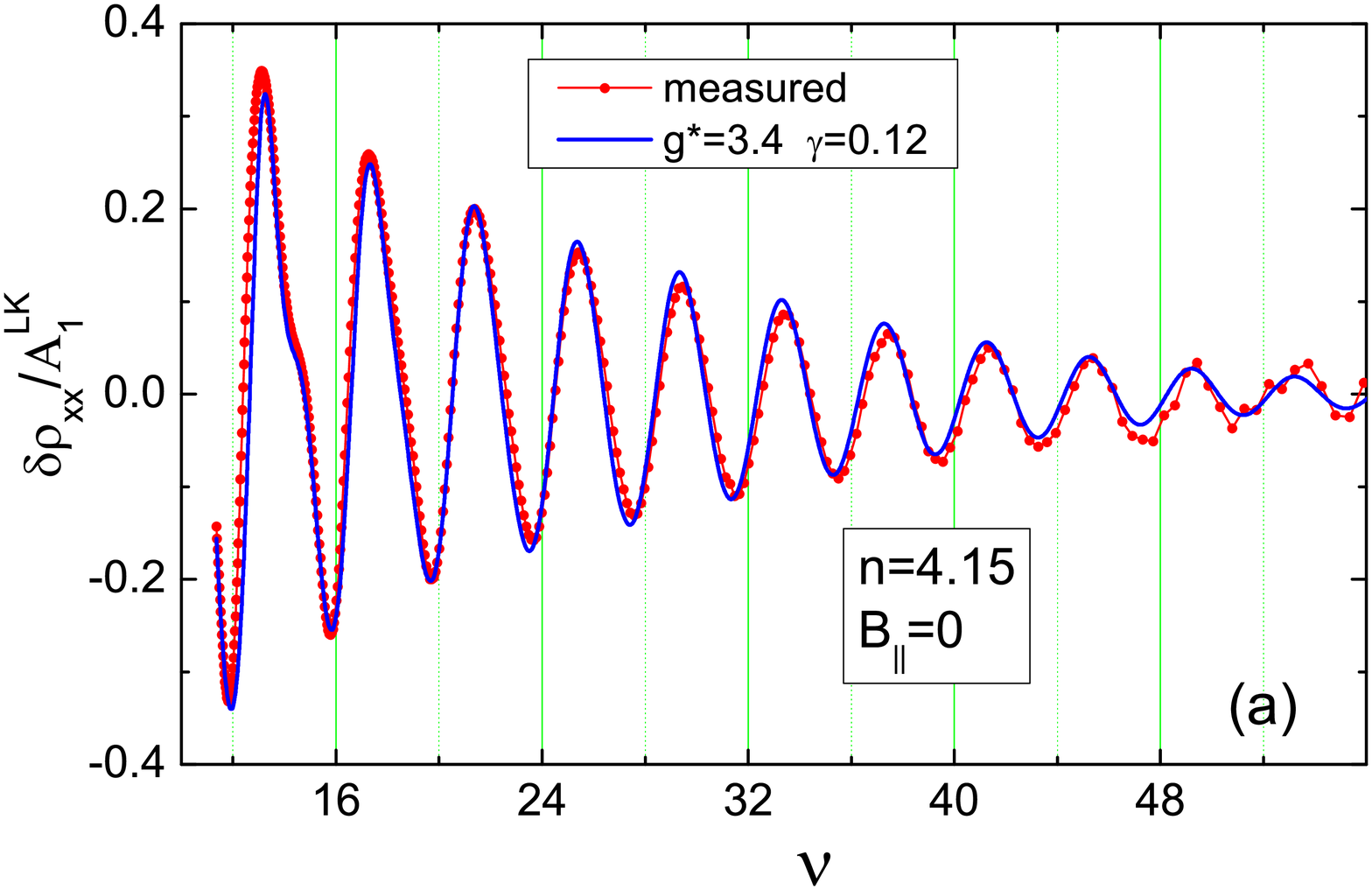}
\includegraphics[width=230pt]{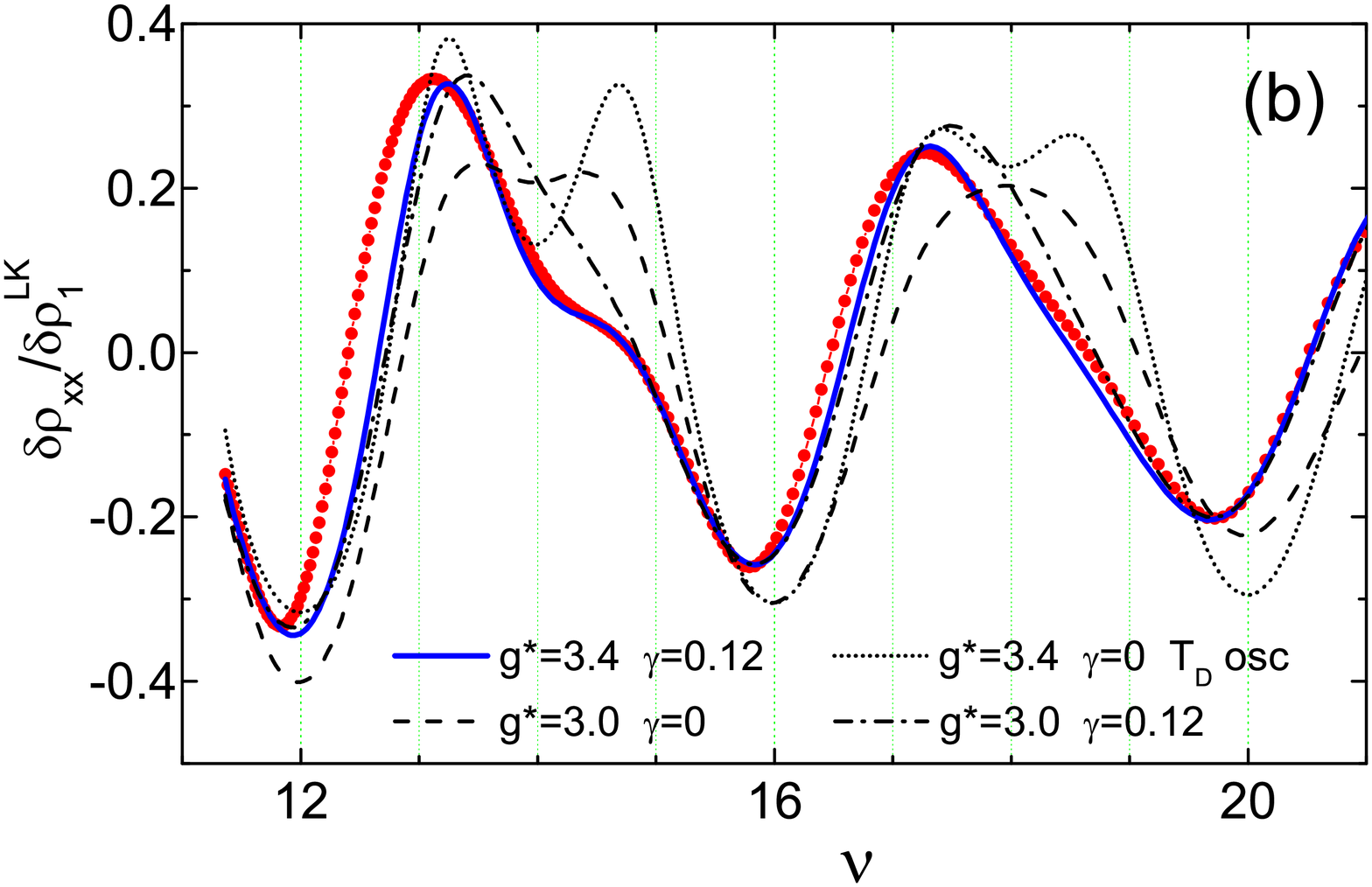}
\caption{(Color online) Normalized SdH oscillations $\delta\rho_{xx}/\delta\rho_1^{LK}$ in the absence of the parallel field vs  filling factor $\nu\propto 1/B_\perp$.
 Carrier density is
 $4.15\times 10^{11}$cm$^{-2}$ ($r_s\approx 4.0$), and
 $T=0.19$K. (a) in the wide range of fields $B_\perp= 0.3 - 1.5$\,T, and (b) oscillation fits corresponding to the sets of parameters shown in the figure.
  }
\label{fig-n4.14_B0}
\end{figure}

\begin{figure}[htp]
\includegraphics[width=230pt]{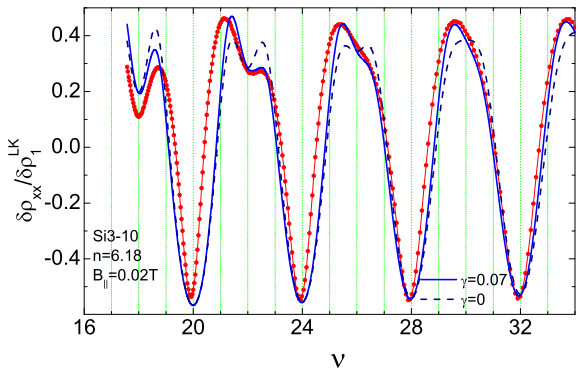}
\caption{(Color online) Normalized SdH oscillations $\delta\rho_{xx}/\delta\rho_1^{LK}$ in the absence of the parallel field vs  filling factor $\nu\propto 1/B_\perp$.
Electron density is $6.18\times 10^{11}$cm$^{-2}$ ($r_s= 3.3$), $g^*=2.9$. Dots are the data, the continuous curve is a fit with $\gamma = +0.07$, and the dashed curve is a fit with $\gamma=0$.
 }
\label{fig-n6.18_B0}
\end{figure}

\begin{figure}[htp]
\includegraphics[width=230pt]{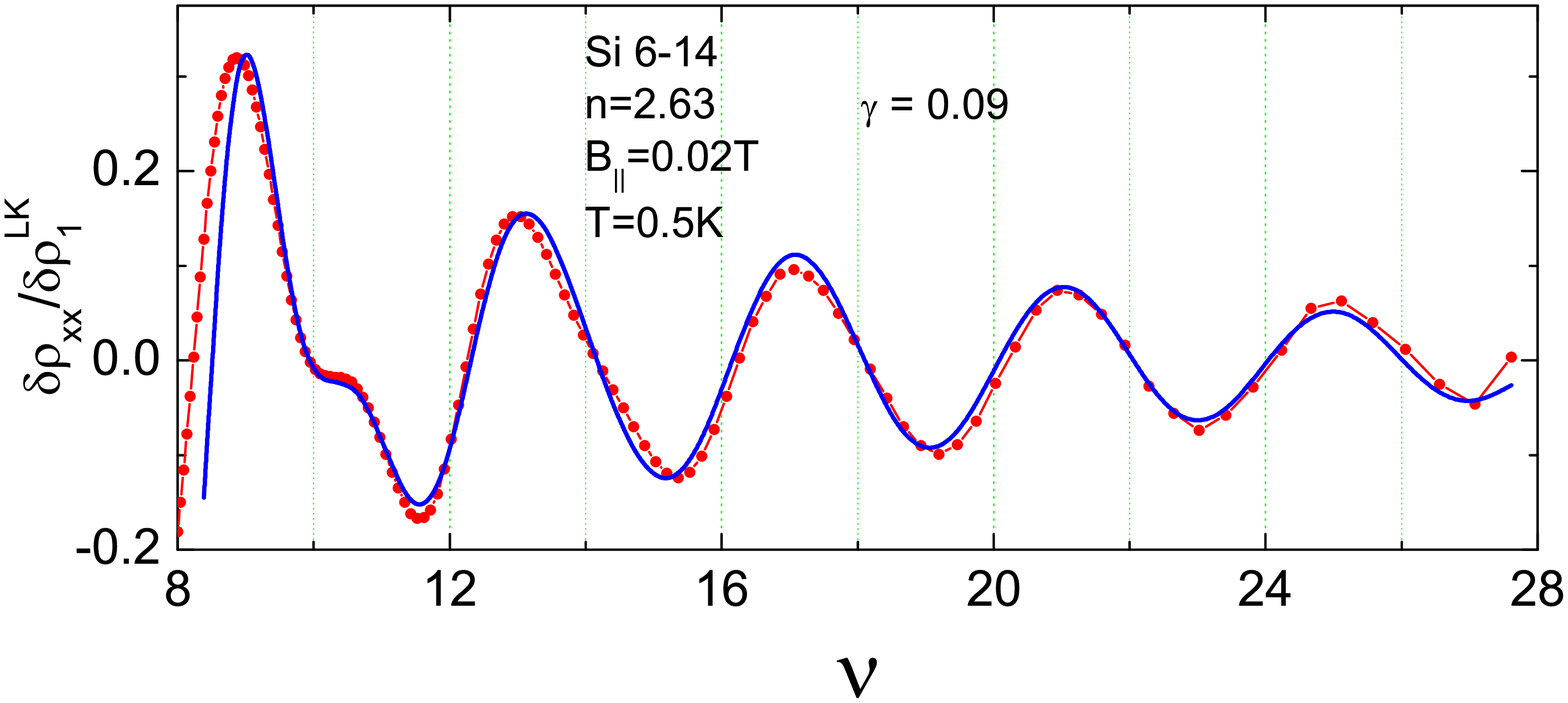}
\caption{(Color online) Normalized SdH oscillations $\delta\rho_{xx}/\delta\rho_1^{LK}$ in the absence of the parallel field vs  filling factor $\nu\propto 1/B_\perp$.
 Electron density is $2.63\times 10^{11}$cm$^{-2}$ ($r_s= 5.07$), $g^*=3.72$ (canonical value 3.24). Dots are the data, the continuous curve is a  fit with $\gamma = +0.09$.
 }
\label{fig-n2.63_B0}
\end{figure}

Beside modeling with an adjustable $\gamma$ value, for comparison, the figures  show examples of modeling with $\gamma=0$ and  with the canonical $g$-factor value.
Figure \ref{fig-n4.14_B0}(b) also shows modeling of the SdH oscillation shape using an oscillatory level broadening according to Eq.~(\ref{TD_ocs}). One can see that the oscillatory $T_D(\nu)$ dependence does not explain the asymmetry of the  oscillation lineshape. Since oscillatory level broadening is irrelevant to both observable features of SdH oscillations, i.e., the extra decay of the amplitude with inverse field and asymmetry of the lineshape, we ignore this effect in the rest of the paper.

\subsection{4.7. Intermediate discussion of the results}
On the basis of the above data analysis we have arrived at several conclusions.

(i) For relatively low filling factors corresponding to the spin gaps $\nu= (4i-2)$ and in zero $B_\parallel$ field,  the $g^*$ factor is enhanced by $\approx 10-15\%$  as compared with its  value previously measured  at $B_\parallel \neq 0$ \cite{pud_gm_2002, klimov_prb_2008}. This enhancement qualitatively agrees   with our earlier measurements of a sharp $\chi^*(B_\parallel)$ field dependence \cite{pud_granada_2004} and with the nonlinearity of the spin magnetization measured in weak  in-plane  fields \cite{reznikov_PRL_2012}.

Alternatively, the $g$-factor  enhancement might also result from the conventional exchange interaction between the adjacent spin levels. It is known that the latter can  increase the  energy  splitting  when the Fermi energy lies in the spin gap  and the difference of spin-subband populations is the largest \cite{macdonald_1986, pud_jetp_1985,  amplitude_enhancement, ando_review}. Although we cannot rule out completely the exchange mechanism,
the interlevel exchange  can hardly explain the enhanced $g^*$ value at Zeeman gaps  $\nu=14, 18,22$ in Figs.~\ref{fig-n4.23_B0}, \ref{fig-n4.14_B0}, and \ref{fig-n6.18_B0}. Indeed, there are two observations at odds with this explanation:
(i) the enhanced $g^*$ value does not decay with $\nu$, being approximately the same for $\nu=14, 18$ and 22 in
Fig.~\ref{fig-n4.23_B0}(b), and for $\nu=14$ and 18 in Fig.~\ref{fig-n4.14_B0}(b), and 
(ii) application of a strong parallel field restores the canonical $g^*$ value (see below).

The same reasoning may be applied to the origin of the extra decay of the oscillation amplitude discussed in section 4.3: it can hardly be related to the exchange interaction enhancement of the oscillations \cite{macdonald_1986, pud_jetp_1985,  amplitude_enhancement}. Indeed,
 as Figs.~\ref{fig-n4.23_B0}(a) and  \ref{fig-n4.14_B0}(a) show, the normalized oscillation amplitude decays with the inverse field down to very weak fields (0.3T) where the amplitude is as small as $\delta\rho/\rho \sim (0.2 - 1)\%$.  At such weak fields $T_D\approx 0.9$K,  $\hbar\omega_c \approx 1.7$K, and
 the effective energy gap $\hbar\omega_c-\frac{1}{2}g^*\mu_B B \approx 1.4$K; hence the neighboring levels overlap heavily and the exchange enhancement is expected to be negligible.

(ii) There is an unexpected difference in scattering rates between the two spin subbands,  which we described by the skew factor $\gamma$, namely, $T_D \propto 1/\tau_q$ is about 20-36\% smaller  in the majority subband, whereas the
spin polarization $\zeta=(n_\uparrow -n_\downarrow)/n$ does not exceed 5\% \cite{note_zeta}. The skew factor, as was mentioned above,
is cooldown dependent (compare Figs.~\ref{fig-n4.23_B0} and \ref{fig-n4.14_B0}).

In the absence of $B_\parallel$ field, the Zeeman splitting in weak fields is relatively small, and the respective
spin dips at $\nu=(4\times i -2)$ can be observed only at the lowest temperatures (they quickly disappear
as temperature increases). For this reason it is difficult to probe the temperature dependence of  $\gamma$ and to clarify its origin. We have solved this problem by performing measurements in nonzero $B_\parallel$,
where the Zeeman splitting is enhanced.
The corresponding  field and temperature dependencies of the skew factor will be discussed below.

(iii)  The field dependence of the oscillation amplitude deviates from the conventional LK-type dependence  $\sim\exp[-2\pi^2 k(T+T_D)/\hbar\omega_c]$, Eq.~(\ref{SdHbyLK}). There is an extra damping of the SdH amplitude with inverse $B_\perp$ field, which we described by the empirical field-dependent $T_D=T_{D0}(1+d_1/B_\perp)$. The extra damping  $d_1$ is  much larger than the interaction correction \cite{adamov_prb_2006}. It is not caused by the interlevel exchange interaction,
hypothetic electron overheating, possible amplitude calibration errors, and possible inhomogeneity of the carrier density over the sample area.

The extra damping factor $d_1$ decreases and changes sign  with temperature  (compare Figs.~\ref{fig-n10_B0} and \ref{fig-n10_B0_T0.5}) and  with $B_\parallel$ field (for more detail, see below). Obviously, such behavior is inconsistent with the interaction correction Eq.~(\ref{reduced LK+gornyi}), which cannot change sign.
Two of the fitting parameters, $d_1$  and $\gamma$, are sample dependent  and  also depend on the temperature and $B_\parallel$ field; $\gamma$, additionally, depends on   the $B_\perp$ field. In the absence of $B_\parallel$  we were able to
reliably disentangle    $d_1$ and $\gamma$ only within a narrow  temperature range 0.1--0.4K.
Their temperature dependencies over a wider temperature range $T= 0.1 - 1$\,K have been measured in the presence of $B_\parallel$ field and will be discussed in the next section.

(iv) Even though the  MO
amplitude is small over the whole range of fields, for the strongest fields $B_\perp \approx 1.3$\,T (i.e. the lowest filling factors $\nu<10$) the MO
lineshape starts deviating from that
calculated  using Eq.~(\ref{SdHbyLK})  [see, e.g., Figs.~\ref{fig-n4.23_B0}, \ref{fig-n4.14_B0}, \ref{fig-n2.63_B0}, and \ref{fig-two examples fitting}(b)].
We attribute these high-field deviations to the interlevel exchange interaction that causes oscillatory level splitting  and oscillatory broadening \cite{macdonald_1986, pud_jetp_1985,  amplitude_enhancement, ando_review}.  This effect is beyond the scope of this paper and we omit the strongest field data in our analysis.

\section{5.~Oscillations in nonzero $B_\parallel$ field. Average characteristics of the partially spin polarized  2DE system }
\subsection{5.1. Brief overview of the $B_\parallel$ field effect on the magnetooscillations}
In the presence of the parallel field $B_\parallel$, the SdH oscillations exhibit beatings; to make the origin of beating  more transparent,
we modify Eq.~(\ref{SdHbyLK}):
\begin{equation}
\frac{\delta\rho_{xx}}{\rho_0} = \sum_i 2\left(a\downarrow^{LK}_i + a\uparrow^{LK}_i\right) \frac{2\pi^2 i k_BT/\hbar \omega_c}{\sinh\left(\frac{2\pi^2 i
k_BT}{\hbar \omega_c}\right)}
\end{equation}
where
\begin{eqnarray}
a(\downarrow,\uparrow)^{LK}_i = \exp\left(-\frac{2\pi^2 i k_B
T_{D\downarrow,\uparrow}}{\hbar \omega_c} \right) \times  \nonumber \\
\times \cos\left\{ i\pi\left[ \frac{\hbar \pi n c}{eB_\perp}\left( 1\pm \frac{g\mu_B B_{\rm total}}{2E_F} \right)-1 \right]\right\} Z_i^v
\end{eqnarray}
with $B_{\rm total}=\sqrt{B_\perp^2+B_\parallel^2}$. One can see that when $B_\parallel \neq 0$, the two oscillatory
patterns interfere causing beatings.

Typical traces of the SdH oscillations in the presence of $B_\parallel$ field  are shown in Figs.~\ref{fig-n9.95_B1.5}, and \ref{fig-n3.76_Bpar}.
Being normalized to the calculated amplitude of the
first harmonic $\delta\rho_1^{LK}$, Eq.~(\ref{SdHbyLK}), the  oscillations exhibit a well pronounced beating pattern [see Figs.~\ref{fig-n9.95_B1.5}(b) and \ref{fig-two examples fitting}].
The oscillations envelope and phase, as well as the node position of beatings
carry information on the $g$-factor value, and on the relative amplitude of the
oscillatory patterns
generated by two spin subbands; the beating pattern is the subject of the analysis in this section.

\begin{figure}[ht]
\includegraphics[width=230pt]{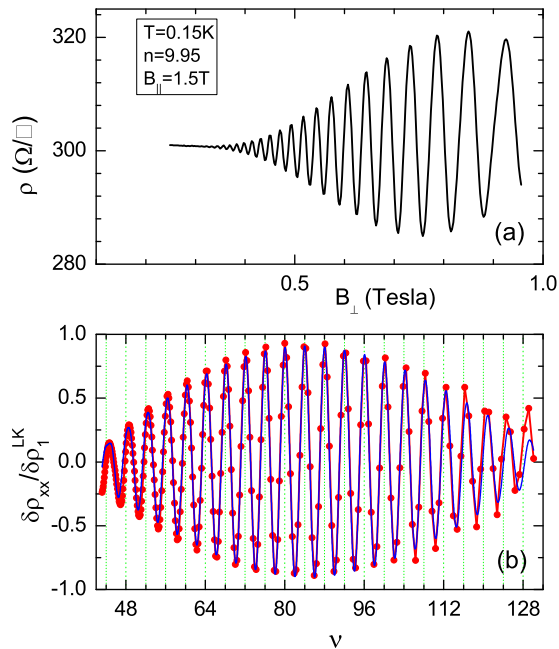}
\caption{(Color online) SdH oscillations in the presence of the parallel field: (a) $\rho(B_\perp)$ data for $B_\parallel =1.5$T and for almost the same density as that in Fig.~\ref{fig-n10_B0}. The temperature is $T=0.15$\,K. (b) Normalized oscillations $\delta\rho/\delta \rho_1^{LK}$ (dots) and their fitting  (line) with Eq.~(\ref{SdHbyLK}) using $T_D=0.87$K, $d_1=0$, $\gamma=0$, and $g^*=2.65$.}
\label{fig-n9.95_B1.5}
\end{figure}

\begin{figure}[ht]
\includegraphics[width=210pt]{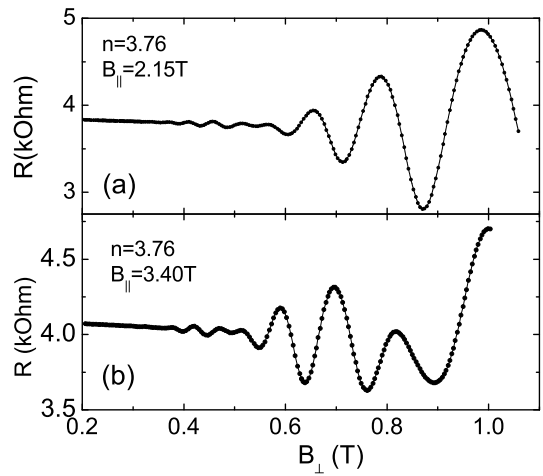}
\caption{Examples of SdH oscillations in the presence of the parallel field: (a) for $B_\parallel =2.15$\,T,  and (b) for 3.4\,T.
The temperature is $T=0.1$\,K. Carrier density is in units of $10^{11}$cm$^{-2}$.}
\label{fig-n3.76_Bpar}
\end{figure}

Application of the in-plane field  unexpectedly  produces
several remarkable effects: \\
 (i) the extra damping of oscillations [$d_1$ in Eq.~(\ref{eq:empiric_TD(B)})]  decreases significantly or vanishes,\\
 (ii) the skew factor $\gamma$ is significantly reduced, and \\
 (iii) $g^*$ factor   regains its canonical value \cite{note-renormalized}.

 Figures \ref{fig-n9.95_B1.5} and \ref{fig-two examples fitting} illustrate that oscillations are now much better described by
 Eq.~(\ref{SdHbyLK})  and that their normalized magnitude remains independent of $B_\perp$ field.  We shall discuss this
 remarkable observation later, and now we return to the ``conventional'' behavior.
  Thanks to the high accuracy of the fitting, $< 1\%$, demonstrated in Figs.~\ref{fig-n9.95_B1.5}(b) and \ref{fig-two examples fitting},
  the fitting parameters are determined with rather high precision.
  The high accuracy of the parameters extraction in the presence of $B_\parallel$ field allows us
   to perform  a comparison of oscillations with the interaction correction theory.

\begin{figure}[ht]
\includegraphics[width=230pt]{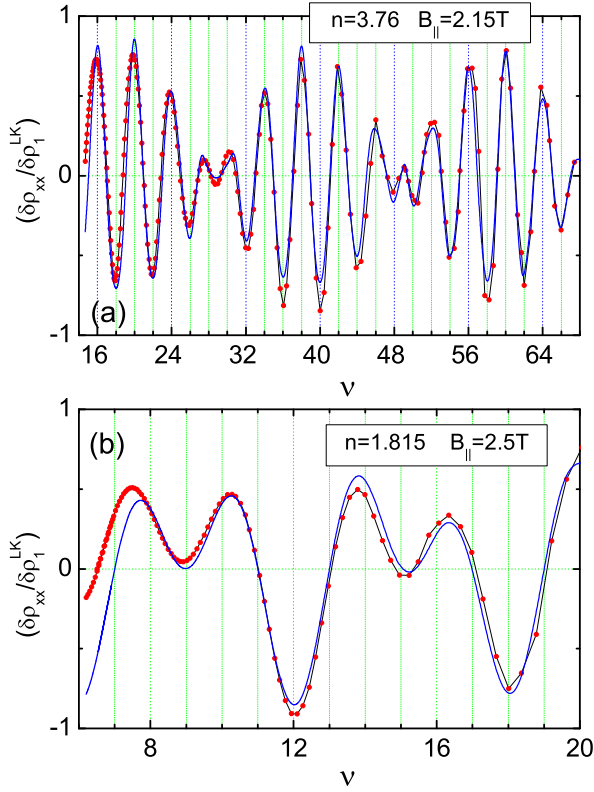}
\caption{(Color online) Examples of
fitting  with Eq.~(1): (a) $n=3.76$, $B_{\parallel}=2.15$\,T,  and (b) $n=1.815$, $B_{\parallel}=2.5$\,T.
The temperature is $T=0.1$\,K, and the density is in units of $10^{11}$cm$^{-2}$. Dots are the normalized data $\delta \rho_{xx}/\rho_1^{LK}$,  curves are the fits with
adjustable  $T_D$, $d_1$,  $\gamma$;  $m^*(n)$ and $g^*(n)$ are the canonical values from Refs.~\cite{klimov_prb_2008, pud_gm_2002}.
 $T_{D0}=0.77$\,K, $\gamma=0.8\%$,  and $d_1=0.07$ for panel (a). $T_{D0}=0.87$K, $\gamma=4$\%, and $d_1=0.17$ for panel (b).
Note that all $\rho_{xx}$-minima on panel (b) correspond to the ``spin gaps'', $\nu=4i-2$ \cite{phase}
}.
\label{fig-two examples fitting}
\end{figure}

\subsection{5.2. Taking interaction into account
}

When comparing the MO amplitude for various $B_\parallel$ fields, we face a dilemma: which conductivity value
should be used for normalization of the MO amplitude
in Eq.~(\ref{SdHbyLK}),
$\sigma_D (B_\parallel=0)$ or $\sigma(T\rightarrow 0, B_\parallel)$?
Neither the LK theory, nor the theory in Refs.~\cite{martin_prb_2003,  adamov_prb_2006}
consider the in-plane  field, and, therefore, do not answer this question directly. In the framework of the interaction correction theory, $\sigma_D$ should not be affected by the temperature and  in-plane field, whereas both $\tau_q$ and $\tau$  are renormalized by interactions causing  temperature- and field-dependencies of the conductivity.
This suggests that the oscillation amplitude in Eq.~(\ref{SdHbyLK})
should be normalized to $\sigma_D$.

On the other hand, it was experimentally established that for
Si-MOSFETs the magnetoresistance in the $B_\parallel$ field is not entirely  described by the interaction corrections:
it is also strongly
dependent  on the magnetic field contribution to the mobility (the so-called
``magnetic field driven disorder'' \cite{disorder, alexkun_2013}). In this case,  if
the disorder is altered by the parallel field \cite{disorder, Das_B-driven_disorder_2005}, the scattering time would be field-dependent and, hence, the oscillation magnitude should be normalized to the field-modified $\sigma_D(B_\parallel)$. The latter value in the ballistic regime (i.e., ignoring logarithmic corrections) may be found as   $\sigma(T\rightarrow 0, B_\parallel)$ \cite{pud_prl_2003}.

The relative share of the two contributions to the magnetoresistance, the interaction corrections and magnetic field driven disorder,   depends on the particular sample. In view of this uncertainty, and
in the spirit of the theory with which we compare our data,
throughout the paper we normalize the MO amplitude to $\sigma_D(B=0)$  even in the presence of $B_\parallel$ field; we have verified that the normalization  to $\sigma_D$ or $\sigma(B_\parallel, T=0)$ results in a minor quantitative difference and does not affect our qualitative results and conclusions.

\section{6.~  Individual renormalizations  in each
subband of the partially polarized 2D electron system}

We now analyze the line shape of SdH oscillations  in order to determine the quantum scattering time in each spin subband
as a  function of the in-plane field and temperature. The emerging spin-splitting causes  dips at the maxima of $\rho_{xx}$ oscillations
at $\nu=4i-2$ which grow with $B_\perp$   (i.e., as $\nu$ decreases). Figures \ref{fig-n6.18_B0} and \ref{fig-6.18_Bpar} show  the development of the spin structure in SdH oscillations at relatively high densities
with an increasing in-plane field.

\subsection{6.1. Background of the data analysis}
In principle,  one can determine all the  renormalized
 parameters of the  electron system, $m^*$, $T_D$, $g^*$, and the  skew parameter $\gamma$,  by fitting the interference pattern of quantum oscillations with Eqs.~(1) -(5). However,  the first two parameters that control the  damping of the average MO amplitude are strongly correlated over the  experimental ranges of temperatures and fields. For this reason, their product $m^*T_D$ can be determined much more reliably. The $g^*$ factor  controls the characteristic fields where the dips appear due to the emerging spin splitting, as well as the nodes of beats. In its turn,  $\gamma$ controls mainly the asymmetry and magnitude of the oscillation pattern near the nodes and does not correlate with $T_D$.

 The  parameters $g^*$ and $\gamma$ are almost  uncorrelated, and this fact enables us to disentangle them with a reasonable accuracy,  $\sim 1.5\%$ for  $g^*$ factor, and up to $\pm 0.002 $ for $\gamma$. We have taken into account the  $\gamma(B_\perp,T)$ dependencies by analyzing the evolution of oscillations  with temperature and  $B_\parallel$ field  for several fixed values of  $B_\perp$ field at which we could extract $\gamma$ from the interference pattern.

 In order to decrease the number of fitting parameters, we analyzed the data using the following scheme. Initially, the beating pattern of oscillations versus $B_\perp$ field  was fitted using  Eq.~(\ref{SdHbyLK}) for a given density, $B_\parallel$ field, and temperature. The  initial $m^*(n)$ and $g^*(n)$  values were calculated using the polynomial approximation of the experimental data \cite{note-renormalized}. The actual $g^*$ value was further fine tuned by fitting  the line shape of the nodes of oscillations and emerging spin gaps. As we mentioned above,   for weak or zero $B_\parallel$, the canonical $g^*$ value had to be increased by about 10\%. This approach leaves us with only  three adjustable parameters:  $T_{D0}$, its field dependence ($d_1$), and $\gamma$. The dependencies of these parameters on $B_\parallel$ field (see below) represent one of our main results.

 As we have already mentioned, the analysis is simplified essentially in the presence of a strong   field $B_\parallel \gtrsim k_BT/2\mu_B $. In this case
 the oscillations damping is reasonably well
 described with a $B_\perp$-{\em field independent} $T_D$ value (i.e.  $d_1$ may be neglected), and   $g^*$ factor regains its  canonical value.

 At the next step, for fixed $B_\parallel$ and  $B_\perp$ fields, we analyzed the  product $m^* T_D$  as a function of temperature. These results will be compared  with the  theory Eqs.~(\ref{eq:gornyi}) and (\ref{reduced LK+gornyi}).
Finally, we found that the extracted $g^*$ and $\gamma$ values vary slightly with perpendicular field; this dependence will be also discussed below.

We fitted the oscillations
with  Eqs.~(\ref{SdHbyLK}) and \,(\ref{eq:gornyi})  using  individual products $m^*T_{D}$ for
each  spin subband,
$$
(m^*T_{D\uparrow \downarrow}) = m^*_0 T_{D0}(1 \pm \gamma)
$$
where  the  average (``zero-field'') mass $m^*_0(n)$  corresponds to the canonical value
~\cite{pud_gm_2002,note-renormalized}.
The sign of $\gamma$ in the above equation is chosen in line with our intuitive expectations and the RPA results:
for the majority ($\uparrow$) subband, the carrier density is larger, the interaction strength $r_s $ is weaker, and screening is stronger; both latter factors are expected to lower the $m^* T_D$ value.

Strictly speaking, relying solely on the experimental data, it is difficult to determine to which of the two parameters
(either $m^*$, or $T_D$) the skew factor $\gamma$  is related to. However, the uppermost RPA estimate of the skewness in $m^*$, Eq.~(\ref{RPA estimate_of skewness}),  is much smaller than the observed  skewness in
the MO amplitude in a purely perpendicular field. As will be shown below, the observed skewness  tends to decrease with $B_\parallel$ field and, hence, with spin polarization, in direct contradiction with the RPA result.
Moreover, according to recent calculations \cite{maslov_prl_2005},    the  difference between  $m^*_{\uparrow}$ and $m^*_{\downarrow}$  for a large-degeneracy (bosonic) 2D gas should be small and its dependence on the spin polarization should be very weak.
Taking this into account, as well as the experimental evidence that
$T_D$ is noticeably dependent on $B_\parallel $ (see below), we have chosen  to associate $\gamma$ with skewness in $T_{D\downarrow,\uparrow}$ in the presence of a strong polarizing $B_\parallel$ field, in the same way as above in Eq.~(\ref{skew_factor})
for a purely perpendicular field.

Figure \ref{fig-6.18_Bpar} shows that
$\gamma$ can be found by
fitting the beats of the oscillations with rather high precision, typically $\pm 0.2\%$. The most sensitive to the $\gamma$ value are the MO amplitude
and phase near the nodes. Additional  information on $\gamma$, even at zero parallel field, comes from emerging spin splittings in the vicinity of
 $\nu=14, 18$, and 22 in Figs.~\ref{fig-n4.23_B0}(b), \ref{fig-n4.14_B0}(b), and
\ref{fig-n2.63_B0}. Correspondingly,
$\gamma$ was determined as a function of $B_\parallel$ at several $B_\perp$ values.

\begin{figure}
\includegraphics[width=240pt]{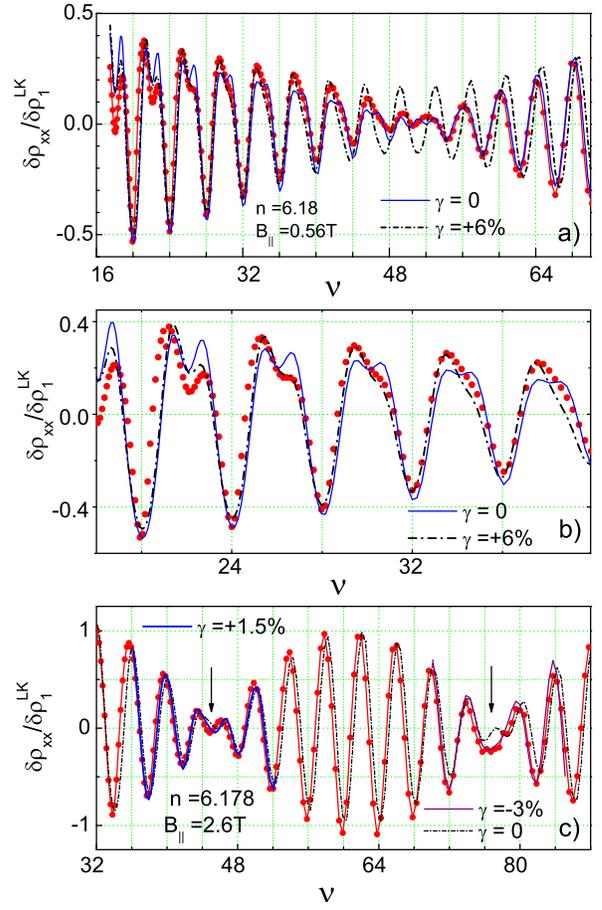}
\caption{(Color online) Examples of fitting SdH oscillations for sample Si3-10 at  $n=6.178\times 10^{11}$cm$^{-2}$: (a)
$B_\parallel=0.56$T, $T_{D0}=0.43$K, and $T= 0.15$K.  (b) blow-up of the low field range $\nu=16$ -- 48 of the same data.
(c) $B_{||}=2.6$T, and $T_{D0}=0.45$K. Dots show the normalized data, the continuous and dash-dotted curves
show  fittings with  various $\gamma$ values. Vertical arrows point at two nodes.
}
\label{fig-6.18_Bpar}
\end{figure}

\begin{figure}
\includegraphics[width=240pt]{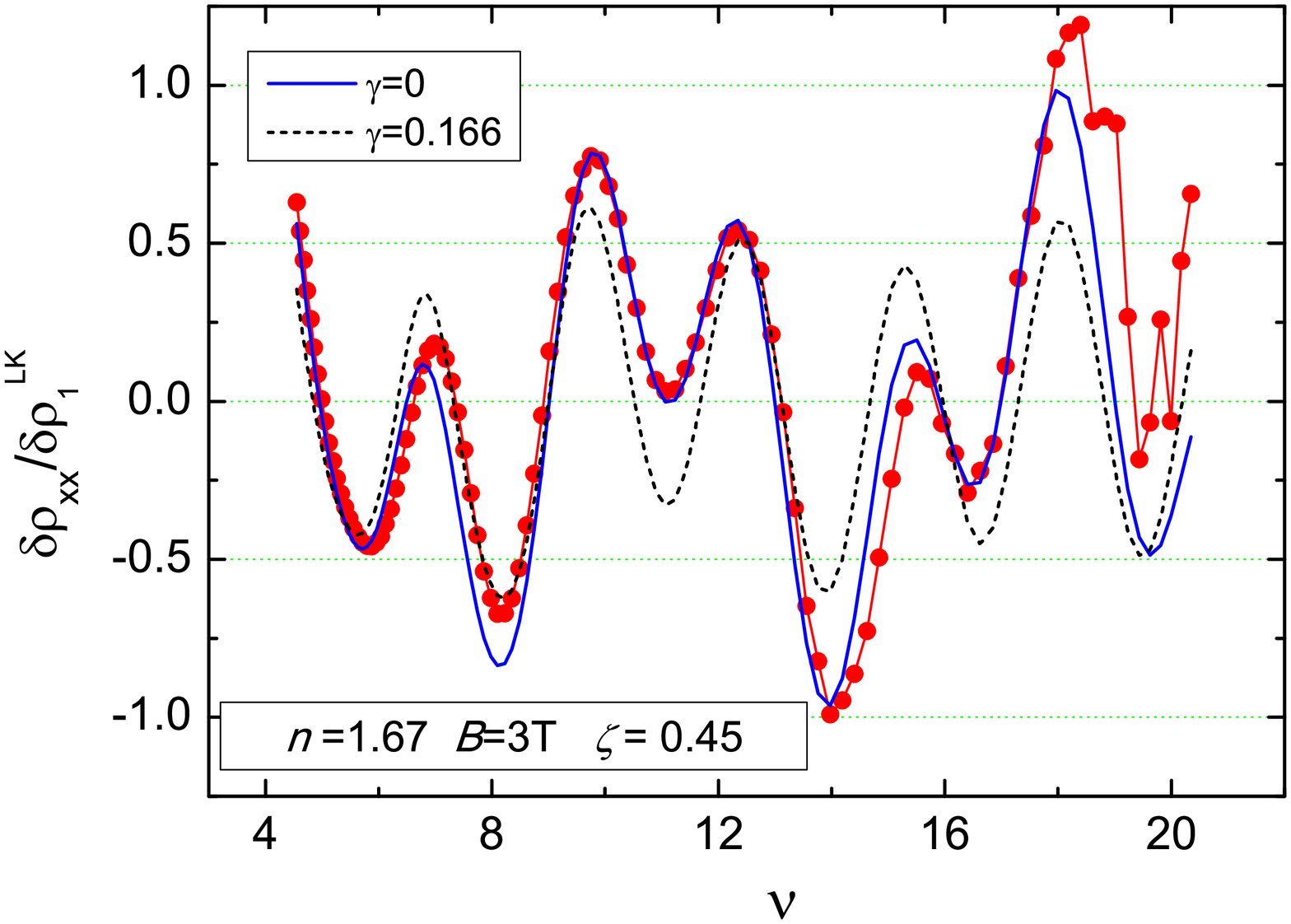}
\includegraphics[width=240pt]{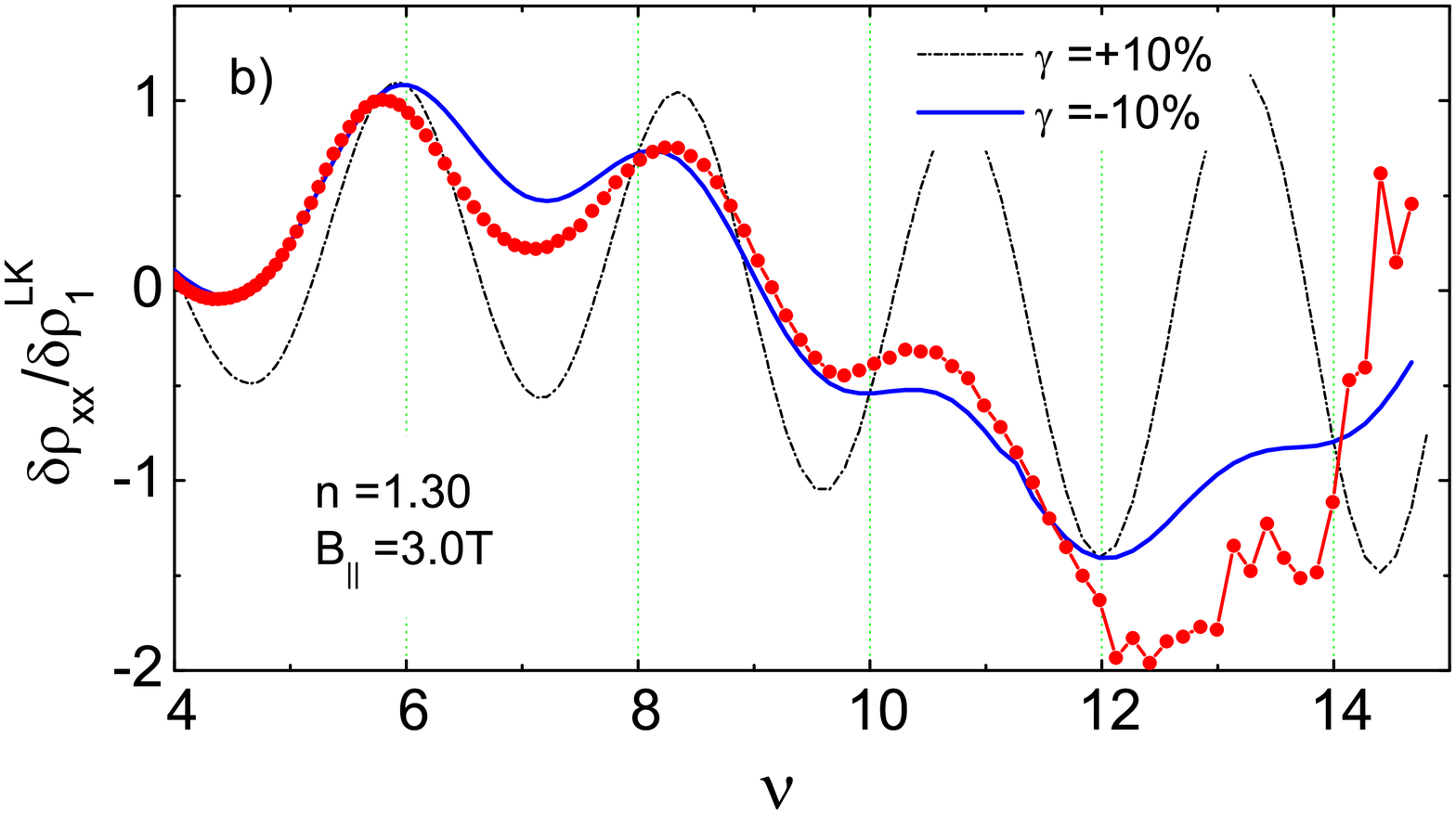}
\caption{(Color online) Examples of fitting SdH oscillations for sample Si3-10 at $B_{||}=3.0$T:  for $n=1.67 \times 10^{11}$cm$^{-2}$,  $\zeta=45\%$ (top) \cite{note_zeta},
and $n=1.3 \times 10^{11}$cm$^{-2}$, $\zeta=66\%$), $T_{D0}=0.67$K (bottom).  Dots show the normalized data,  continuous and dashed curves show  fittings with various $\gamma$ values.
}
\label{fig-1.7_B3}
\end{figure}

\subsection{6.2. Variation of  $\gamma$ and $d_1$ with field and temperature}

In  the absence of the parallel field [Fig.~\ref{fig-n6.18_B0}(a)],
$\gamma$ has the anticipated  sign; its value ($\sim +7$\% in strong field $B_\perp = 1.4$\,T)  is of the order of the polarization degree
$\zeta= \sim 4\%$ \cite{note_zeta}.
Figures \ref{fig-6.18_Bpar}(a) and \ref{fig-6.18_Bpar}(b) show that
in the presence  of $B_\parallel= 0.56$\,T,  $\gamma$ changes  from 6\% (at $\nu =20 - 40$) to almost zero in weak perpendicular fields (at $\nu\approx 48 - 54$, $B_\perp \approx 0.5$\,T).
Closer inspection of  Fig.~\ref{fig-6.18_Bpar}(a) reveals that this change occurs rather
abruptly, between $\nu= 42$ and 49.
We note that the Zeeman energy at the corresponding perpendicular field $B_\perp = 0.6 -0.53$\,T is ten times greater than the temperature $0.15$K and is of the order of the level broadening $T_D$. The role of the latter relation will be discussed below.

In stronger
parallel field $B_\parallel= 2.6$\,T ([Fig.~\ref{fig-6.18_Bpar}(c)] the  $\gamma$  value similarly decreases from 1.5\%  to -3\%  as $B_\perp$ field decreases (i.e., $\nu$ increases from 72 to 80).
And vice versa,  $\gamma$  tends to regain its original value with increasing $B_\perp$ field: at  $B_\perp$ corresponding to $\nu\approx 44$, $\gamma$ increases to +1.5\%.  Note that the polarization degree
varies negligibly, from $\zeta(B_{\rm total}) = 8\%$ to 7.5\%, over the range of fields in Fig.~\ref{fig-6.18_Bpar}(b) and hence is irrelevant to the sign change of $\gamma$. We also note that the oscillation line shape evolution with the field is fully reproducible and reversible.

The negative sign of $\gamma$  implies that the $m^*T_D$
product in the majority spin  subband becomes {\em larger} than that in the minority subband,
a result that obviously contradicts the common sense arguments based on the screening and RPA approaches.
The $\gamma $ decrease with $B_\parallel$ field is observed for the whole explored range of carrier densities,
$ 1.3\times 10^{11} <n <10^{12}$cm$^{-2}$; it  becomes very pronounced at lower densities (Fig.~\ref{fig-1.7_B3}).
In the latter case for $\zeta = 66\%$, $m^*T_D$  for  the majority
 subband is  by 20\% {\em larger} than that for the minority subband.

\begin{figure}
\includegraphics[width=240pt]{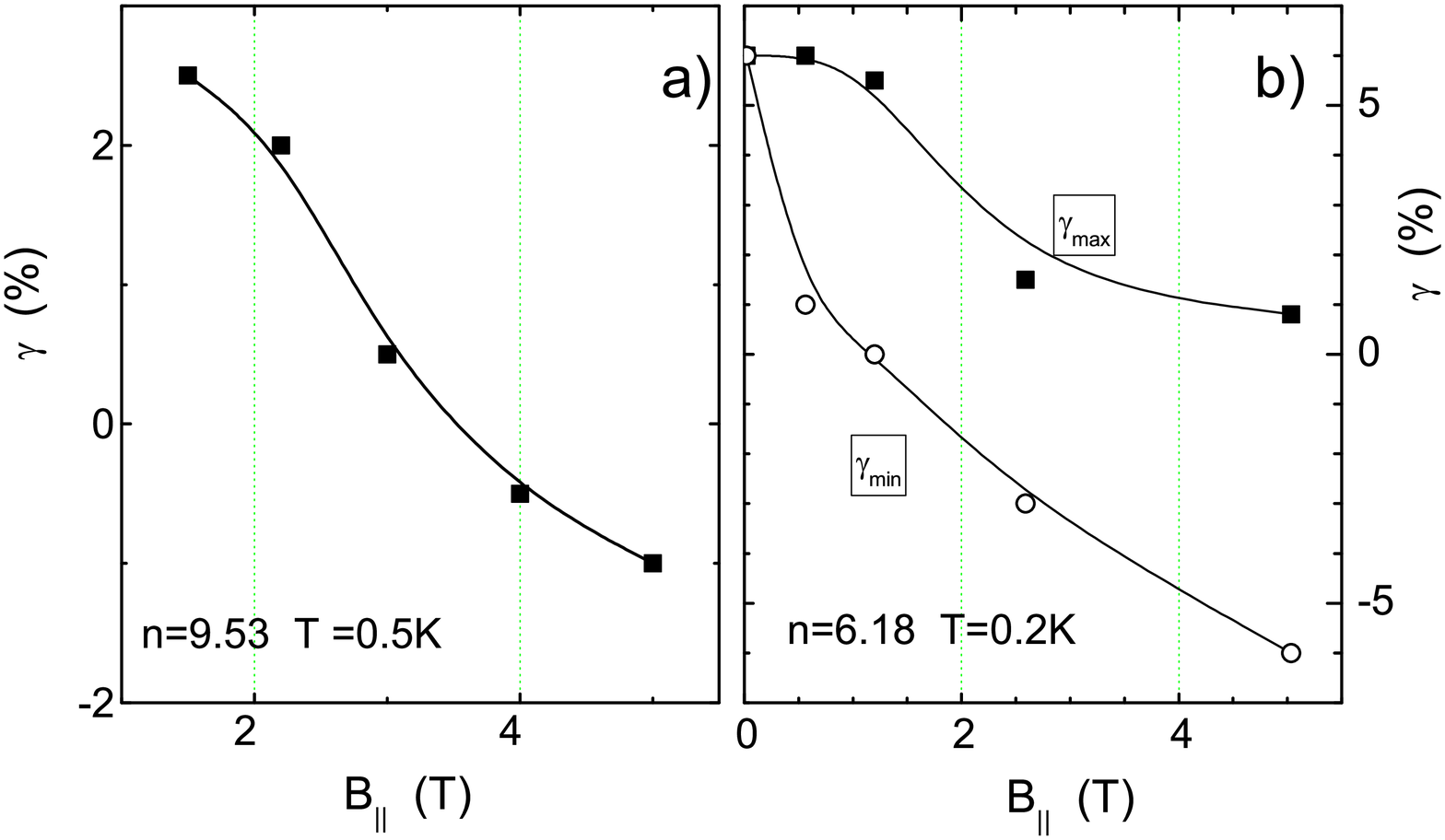}
\includegraphics[width=160pt]{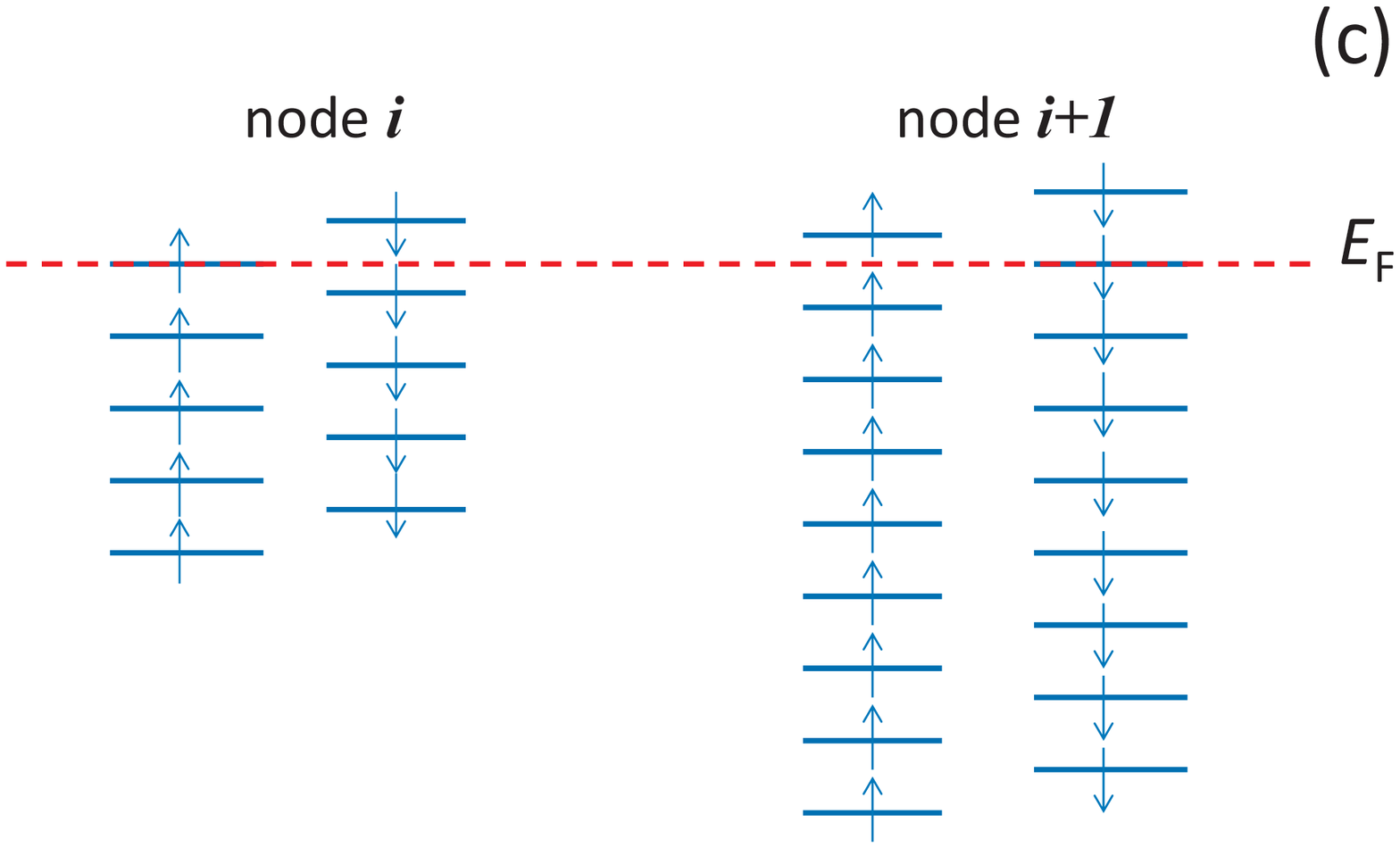}
\caption{(Color online) $B_\parallel$ field dependencies of  the skew factor $\gamma$:
(a) for $n=9.53\times 10^{11}$cm$^{-2}$ extracted for a single node,
(b) for $n=6.178\times 10^{11}$cm$^{-2}$ for two nodes (upper curve for larger $B_\perp$, and lower curve for smaller $B_\perp$).
(c) Schematic diagram of the Landau levels for two spin subbands, in the vicinity of two sequential nodes.
}
\label{gamma_vs_B_parallel}
\end{figure}

A typical variation of the skewness $\gamma$ with in-plane field is plotted in Figs.~\ref{gamma_vs_B_parallel}(a) and \ref{gamma_vs_B_parallel}(b) for two carrier densities. The trend is similar for all densities, but
at lower densities the decrease of  $\gamma$ is more pronounced and  $\gamma$ drops to
more negative  values. For low densities, more than
one node could be observed within the accessible range of $B_\perp$ fields; the respective $\gamma$ values for two nodes  are shown in
Fig.~\ref{gamma_vs_B_parallel}(b). Again, in stronger $B_\perp$ fields (upper curve), the dependence  $\gamma(B_\parallel)$ is weaker.
Obviously, the variation of $\gamma$ values
for the two nodes  [(Fig.~\ref{gamma_vs_B_parallel}(b)] does not correlate with the polarization degree of the 2D system of mobile electrons
$\zeta(B_{\rm total})$  and, thus, cannot be attributed to it.

As we have mentioned above,  $\gamma$ changes with $B_\perp$ field, e.g., between two curves in Fig.~\ref{gamma_vs_B_parallel}(b),
rather abruptly near the node.  This clearly points to the relevance of the interlevel exchange.
The situation  illustrated schematically in Fig.~\ref{gamma_vs_B_parallel}(c) shows that the two subbands are inequivalent in the node vicinity.
 The difference in the population of spin-up and spin-down quasiparticles is the largest near the nodes, and the effective spin gap is expected to be enhanced. Although this mechanism solely cannot explain the variation of $\gamma$ with in-plane field, we conclude  that the interlevel exchange cannot be ignored completely.

\begin{figure}
\includegraphics[width=230pt]{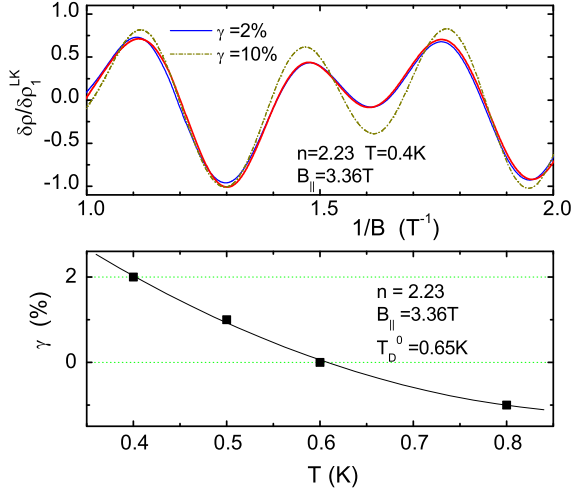}
\caption{(Color online)
(a) SdH oscillations fitting at low densities and  in the presence of strong $B_\parallel$ field. $T=0.4$K
and  $\zeta(B_{\rm total}) \approx$ 35\%.
(b) Typical temperature dependence of $\gamma$. Density  is given in units of
$10^{11}$cm$^{-2}$. Sample Si6-14.
}
\label{fig-gamma_vs_T}
\end{figure}

Figures \ref{fig-gamma_vs_T} and  \ref{fig-d1 and gamma_vs_T} show that the
temperature increase leads to the decrease of $\gamma$ similar to that
caused by the in-plane field (compare Figs.~\ref{gamma_vs_B_parallel}a,b and \ref{fig-gamma_vs_T}b).
The  extra damping parameter $d_1$ also decreases and changes sign as the temperature increases (see Fig.~\ref{fig-d1 and gamma_vs_T}). This behavior is qualitatively
 similar for various carrier densities and $B_\parallel$ values.

\begin{figure}
\includegraphics[width=240pt]{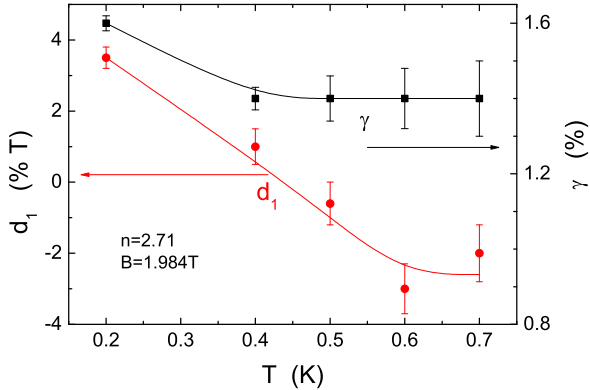}
\caption{(Color online) Typical temperature dependencies of $\gamma$ and $d_1$ measured at fixed $B_\parallel=1.984$T. Density  is given in units of
$10^{11}$cm$^{-2}$. Sample Si6-14.
}
\label{fig-d1 and gamma_vs_T}
\end{figure}

\section{7. Discussion}
\subsection{7.1. On the effective mass equality in the two subbands}
The main  result of our studies  is the approximate equality of the renormalized parameters $(m^*T_{D\downarrow \uparrow})$  in two spin subbands for the spin polarization degree as high as 66\% (see, e.g., Figs.~\ref{fig-6.18_Bpar} and \ref{fig-1.7_B3}). This observation is in line with the results of prior
 measurements \cite{note on mass, shashkin_prl_2003}.
  Most clearly  this equality is illustrated by a nearly vanishing amplitude of beats in the vicinity of nodes of the oscillatory pattern [see Figs.~\ref{fig-two examples fitting}(a) and
\ref{fig-6.18_Bpar}].

Were the electrons interacting predominantly within each spin subband,  the product $(m^*T_{D\uparrow,\downarrow})$ would  have been essentially different with the skew parameter $\gamma$  of the order of polarization $\zeta$.
Our result implies that the exchange  takes place among all the electrons,
 irrelevant to their spins.  This result
 also agrees with a recent theory \cite{maslov_prl_2005} that considers  a large-degeneracy 2D electron gas.

Strictly speaking, an alternative explanation  can also be constructed in a scenario of electrons interacting predominantly within each subband. However, in this case
the mass renormalization  must almost entirely compensate the
scattering rate  renormalization within each subband to provide the same value of $(m^*T_D)$. Taking into  account that the combinations $(m^*T_{D\downarrow, \uparrow})$ remain almost the same for two subbands over ranges of the temperature and parallel field where $T_D$ significantly changes,  this possibility seems  very unlikely. Moreover, in such a scenario of two ``isolated'' spin subsystems, both $m^*$ and $T_D$  should decrease with density , and, hence, their variations cannot compensate each other.

\subsection{7.2. Temperature and field dependencies of magneto-oscillation damping. Comparison with the theory of an interacting 2D systems}

Another goal of our  study of SdH oscillations  in Si-MOSFETs
was to test the theory of quantum oscillations in an interacting 2D electron system.
By measuring the amplitude of quantum oscillations versus the temperature and both field components,
we aimed at verifying the main prediction of the theory 
--   renormalization of the  Dingle temperature $T_D^*(T,B_\perp)$ with the temperature and magnetic field.
This prediction was made for the ballistic interaction regime, which is realized in our high mobility samples at  $T>0.3$K.  To simplify comparison of Eq.~(\ref{reduced LK+gornyi}) with the $T_D(T)$ data, below we discuss the difference $\delta T_D=T_D(T)-T_D(0)$ where the $T_{D0}$ values were estimated by extrapolating $T_D(T)$ to $T=0$.

Figures~\ref{fig-TD_dr_n10_B0_vs_T} show that the monotonic $\delta\rho/\rho$ measured for the same sample in years 2000 and 2006 is highly (within 0.1\%) reproducible. The MO amplitude is less reproducible, it is  cooldown dependent (will be discussed below). For this reason
in the next sections we discuss only the data measured within one cooldown.

\subsection{Comparison with the theory in the absence of $B_\parallel$ field}
 Figures \ref{fig-TD_dr_n10_B0_vs_T} and  \ref{fig-TD_dr_n2.75_B0_vs_T}  show $T_D(T)$
  variations in zero $B_\parallel$ field, for high and low densities, respectively.
  $T_D^*$  trends to grow with $T$, at least in the high temperature range, which is in qualitative agreement  with theory. However, there are several quantitative inconsistencies with  Eq.~(\ref{reduced LK+gornyi}):\\
(i) The experimental slope  $d T_D/dT$ is smaller  (by a factor of 3 -- 10) than the calculated dependence. \\
(ii) The $T_D(T)$ dependence exhibits a minimum at a density-dependent temperature.\\
(iii) The slope  $dT_D/dT$ does not  follow  the predicted $1/B_\perp$ field dependence: it saturates in low $B_\perp$ fields  (see Figs.~\ref{fig-TD_dr_n10_B0_vs_T}, \ref{fig-TD_dr_n2.75_B0_vs_T}, and  \ref{fig-TD_dr_n2.7_B1.98_vs_T}).

According to Eq.~(\ref{reduced LK+gornyi}), the slope $dT_D^*/dT$ depends primarily on $F_0^a$  and the number of triplet terms. Thus, to bring the calculated slope in agreement with our data in Fig.~\ref{fig-TD_dr_n10_B0_vs_T}, we need to assume that either $F_0^a=-0.08$ (instead of  canonical -0.22), or reduce the number of triplets (4.7 instead of 15). Using the same arguments for Fig.~\ref{fig-TD_dr_n2.75_B0_vs_T}, the agreement can be reached by using  $F_0^a=-0.21$ (instead of canonical -0.38), or 6.7 triplets instead of 15. Both assumptions are groundless.

For  all densities  there is an unexpected upturn of $T_D(T)$ at low temperatures;  the temperature of the $T_D(T)$ minimum decreases as the density increases; $\delta T_D(T)$  also turns up at $T<0.2$K  in Fig.~\ref{fig-TD_dr_n10_B0_vs_T}, however we could not quantify it because of a  strong lineshape distortion (caused by the interlevel exchange, as mentioned above)  and therefore do not show the data below $T=0.3$\,K.
The  upturn in $T_D(T)$ is reminiscent of the upturn in $\rho(T)$ \cite{klimov_prb_2008},  where it is caused by intervalley scattering,  valley- and Zeeman splitting, all of which reduce the effective number of triplets \cite{klimov_prb_2008}.
However, the temperature of  $T_D(T)$ minimum is typically  higher than that of $\rho(T)$: the upturn  in $\rho(T)$ was observed below $T\approx 0.2$K for our samples \cite{klimov_prb_2008} at zero field.
Figure \ref{fig-TD_dr_n2.75_B0_vs_T} shows that the $\delta T_D(T)$ minima do not depend on  $B_\perp$, a fact that rules out the effect of the magnetic field enhanced valley splitting.  For this reason, we discuss below another mechanism of the oscillation  damping related to the interface properties.

\begin{figure}
\includegraphics[width=240pt]{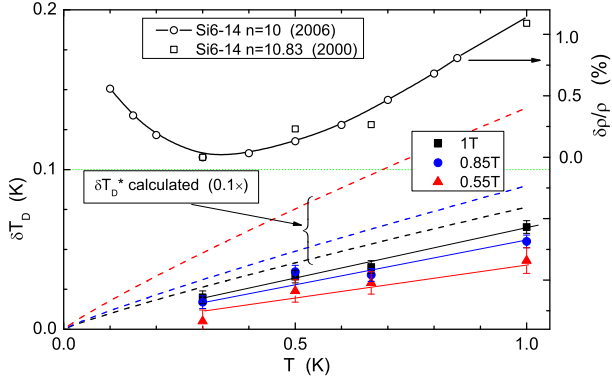}
\caption{(Color online) The dependence $T_D(T)$  measured at  $B_\parallel=0$ and $n=10.83\times 10^{11}$cm$^{-2}$. Full symbols show $\delta T_D=T_D(T)-T_D(0)$ for three $B_\perp$ fields, the connecting solid lines are guide to the eye. $T_D(0) =0.586$K was subtracted to simplify comparison  with the theory. Three dashed  curves show the theoretical $\delta T_D^*(T)$ dependencies (divided by a factor of 10), calculated from Eq.~(\ref{reduced LK+gornyi}) for $B_\perp = 1$, 0.85 and 0.55\,T (from bottom to top). For comparison, the dependence  $\rho(T, B=0)$ is shown with empty symbols. Note a 0.1\%  reproducibility of $\delta\rho/\rho$  measured  for the same sample Si6-14 several years apart.
}
\label{fig-TD_dr_n10_B0_vs_T}
\end{figure}

\begin{figure}
\includegraphics[width=240pt]{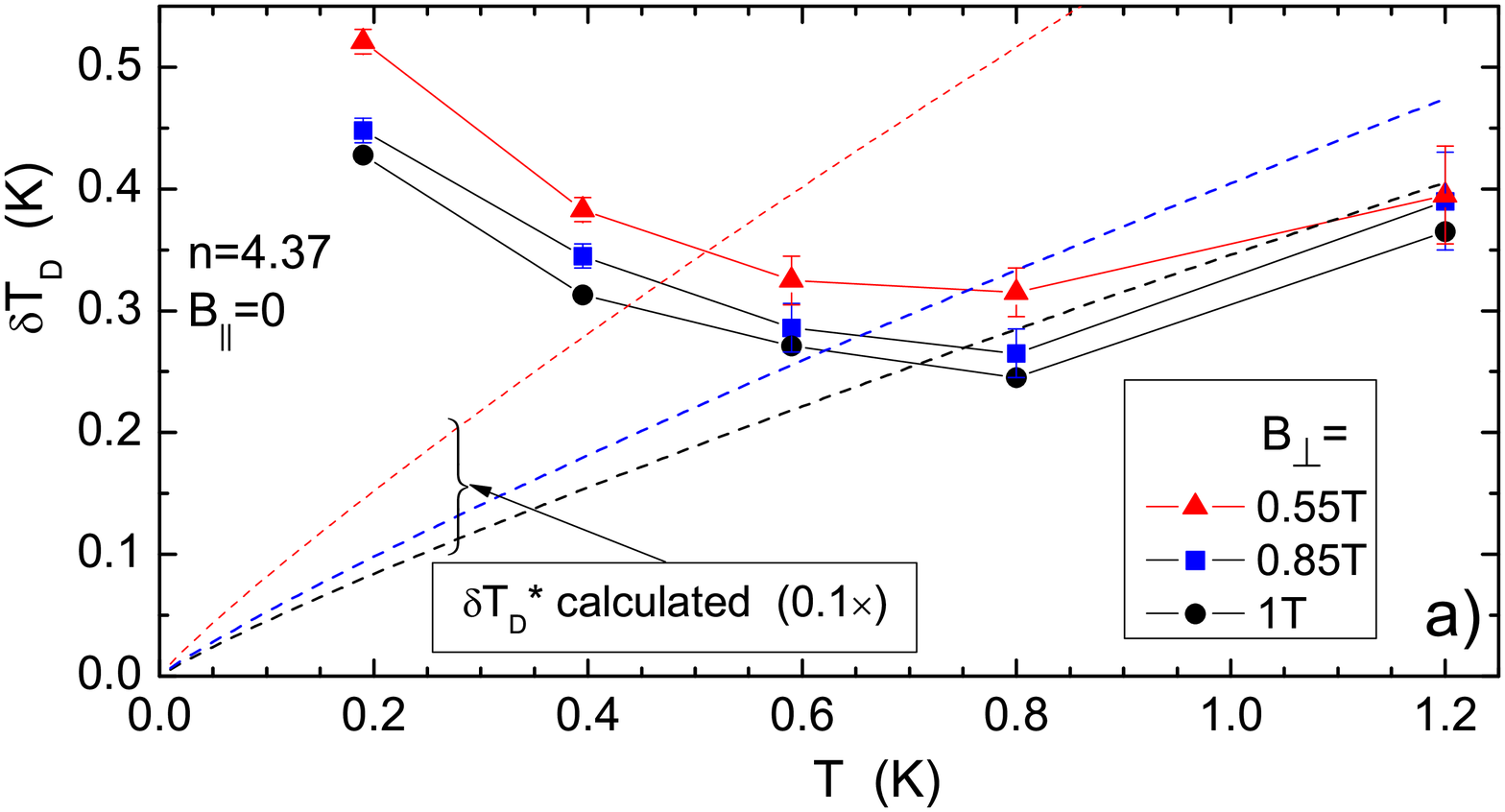}
\includegraphics[width=240pt]{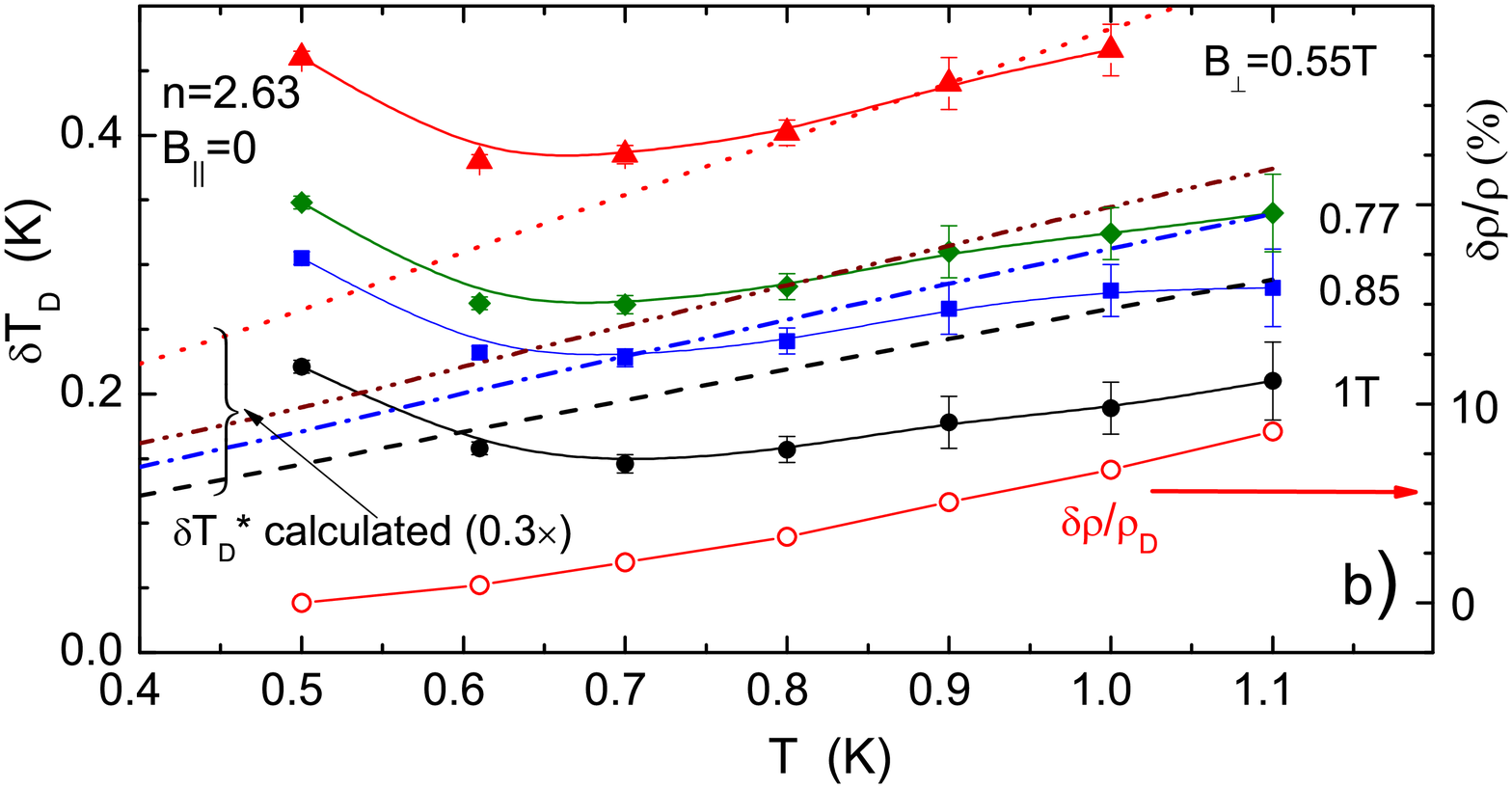}
\caption{(Color online) Comparison with theory of the $T_D(T)$ variations  measured  at  $B_\parallel=0$  for (a) $n=4.37\times 10^{11}$cm$^{-2}$ and (b) for $n=2.63\times 10^{11}$cm$^{-2}$.  Symbols with connecting lines designate in (a)   $\delta T_D =T_D-0.225$\,K values measured at three $B_\perp$ fields, and in (b):  $\delta T_D =T_D-0.3$\,K values measured at four $B_\perp$ fields.  The curves are calculated from Eq.~(\ref{reduced LK+gornyi}) for the same fields (from bottom to top) and reduced by $10\times$ in (a) and by $3\times$  in (b).
For comparison, empty circles show also   $\delta \rho/\rho$ variations  with temperature.
 Sample Si6-14.}
\label{fig-TD_dr_n2.75_B0_vs_T}
\end{figure}

\begin{figure}
\includegraphics[width=240pt]{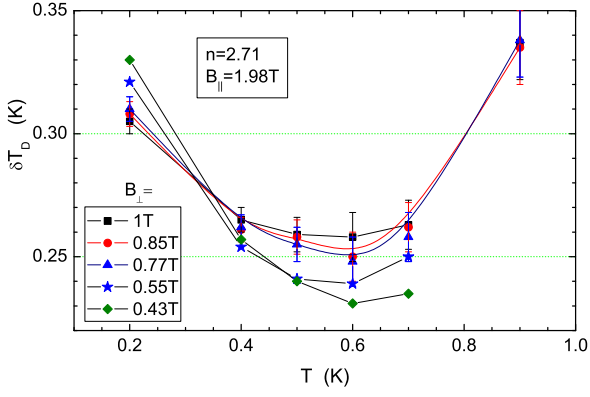}
\caption{(Color online) $T_D$ variations  with temperature. Density $n=2.71\times 10^{11}$cm$^{-2}$ and $B_\parallel=1.984$\,T.
Full symbols are for $\delta T_D=T_D-0.275$K, measured at five different $B_\perp$ fields.  Sample Si6-14.
}
\label{fig-TD_dr_n2.7_B1.98_vs_T}
\end{figure}

\begin{figure}
\includegraphics[width=240pt]{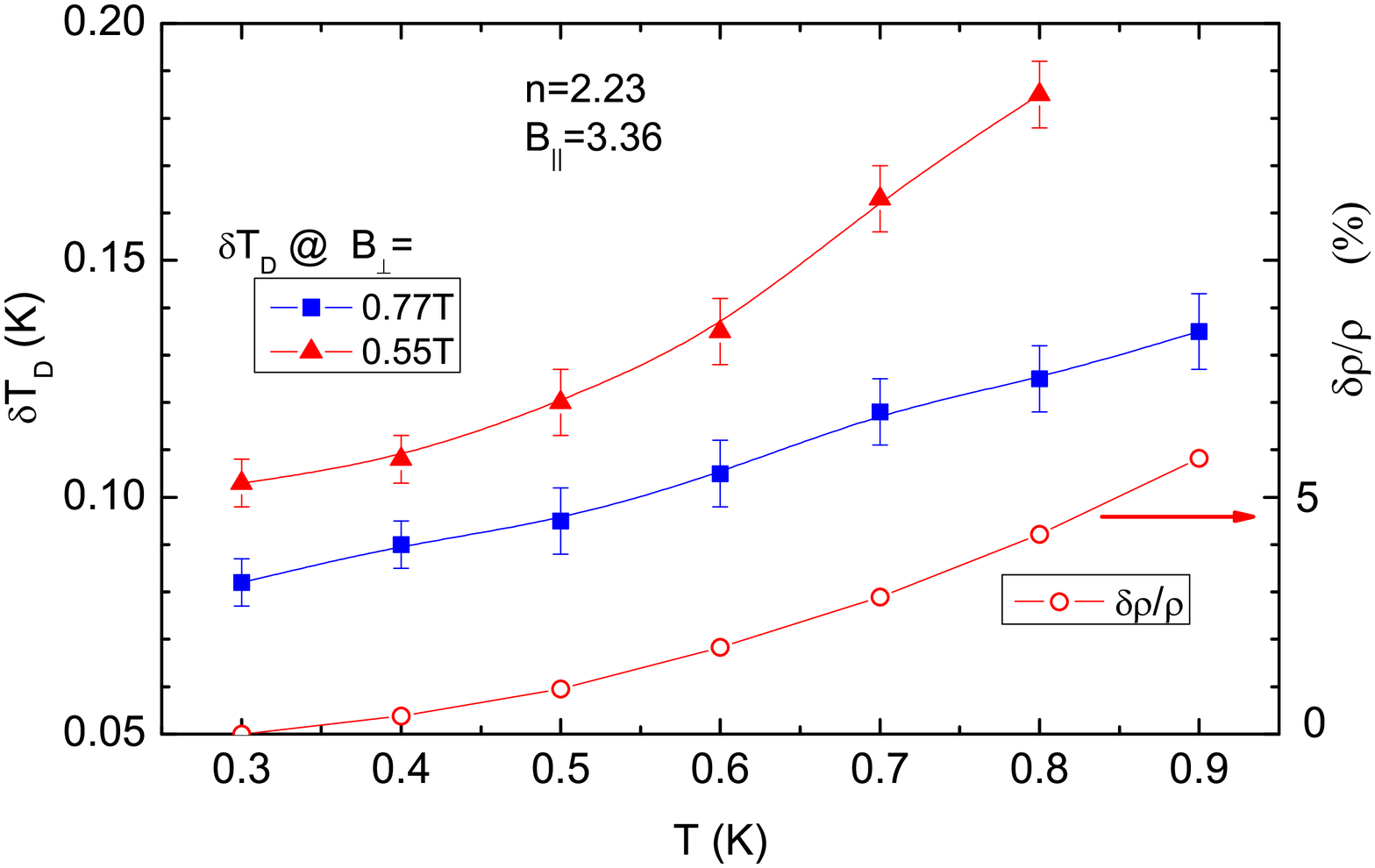}
\caption{(Color online) Full symbols show  $\delta T_D=T_D-0.44$K  variations  with temperature for two values of the $B_\perp$ field. Density $n=2.23\times 10^{11}$cm$^{-2}$, $B_\parallel=3.36$\,T.
Empty circles, for comparison, depict   $\delta\rho/\rho$.  Sample Si6-14.
}
\label{fig-TD_dr_n2.23_B3.36_vs_T}
\end{figure}

\subsection{Comparison with theory in  $B_\parallel \neq 0$ field}
Potentially,  quantitative understanding of the data may be further complicated by the  $B_\perp$-field and the temperature dependencies of $g^*$ and $\gamma$ (see below),  which are beyond the framework of the theory.
Fortunately, for the oscillations measured in non-zero $B_\parallel$, the amplitude of beating antinodes is almost insensitive to $\gamma$ and  $g^*$ values, which allows us  to disentangle $T_D$,  $\gamma$, and $g^*$.

The upturn in $T_D(T)$ usually occurs at  high temperatures for both $B_\parallel=0$ and $\neq 0$ (see Figs.~\ref{fig-TD_dr_n2.75_B0_vs_T}, \ref{fig-TD_dr_n2.7_B1.98_vs_T}, and \ref{fig-TD_dr_n1.77_B3.37_vs_T}).
A closer inspection of the MO data shows that the upturn to a large extent reflects the exchange-enhanced spin gaps (i.e. the quantum Hall ferromagnetism, QHF). We believe that for this reason the upturn is so pronounced in Figs.~\ref{fig-TD_dr_n2.75_B0_vs_T}, \ref{fig-TD_dr_n2.7_B1.98_vs_T},
and \ref{fig-TD_dr_n1.77_B3.37_vs_T}. And vice versa,
in those cases when we were able to trace MO amplitude far away from the corresponding spin gaps (as in Fig.~\ref{fig-TD_dr_n2.23_B3.36_vs_T}), $T_D$ drops monotonically down to the lowest accessible temperature.

The position of the $T_D(T)$ minimum is not affected by $B_\parallel$ (cf. Figs.~\ref{fig-TD_dr_n2.7_B1.98_vs_T} and \ref{fig-TD_dr_n2.75_B0_vs_T}).
This suggests that the $T_D(T)$-minima are related to the physics of a quantizing field rather than to the purely Zeeman splitting.

\begin{figure}
\includegraphics[width=240pt]{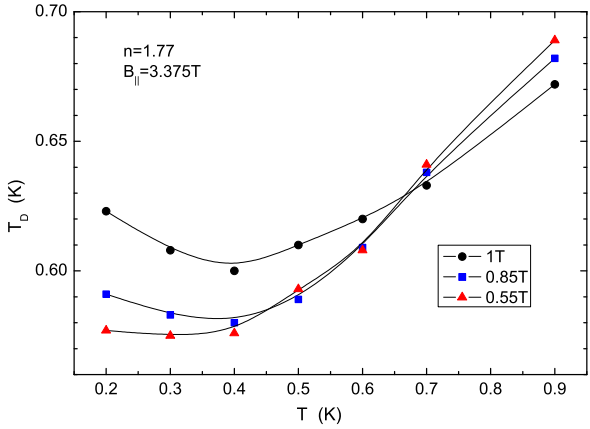}
\caption{(Color online) Full symbols show  $T_D$  variations  with temperature for three values of the $B_\perp$ field. Density $n=1.77\times 10^{11}$cm$^{-2}$ and $B_\parallel=3.375$\,T.
Sample Si6-14.
}
\label{fig-TD_dr_n1.77_B3.37_vs_T}
\end{figure}

Although the theory \cite{martin_prb_2003, adamov_prb_2006} considered  $T_D^*$  renormalization with temperature in the absence of $B_\parallel$, we attempted to test whether  the magnetoresistance $\rho(B_\parallel)$
 and MO amplitude  $T_D^*$ are affected by the same mechanism.

 \begin{figure}
 \includegraphics[width=240pt]{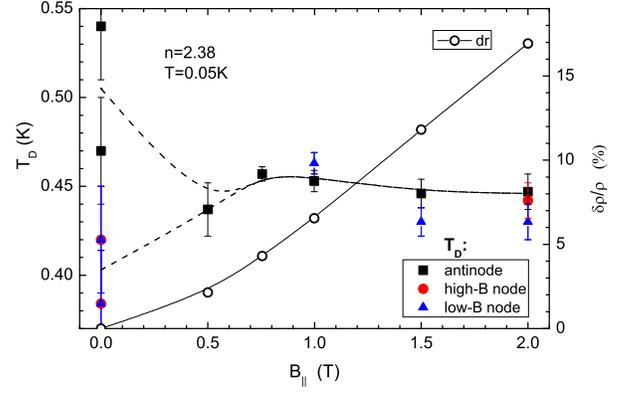}
\caption{(Color online) Variations of $T_D$   with in-plane field. The temperature is $0.05$K and the density is $n=2.38\times 10^{11}$cm$^{-2}$.
Empty circles show $\delta\rho/\rho$, full symbols show  $T_D$ measured at  different $B_\perp$ fields,  corresponding to the two nodes  and one antinode. Sample Si6-14.
}
\label{fig-n2.38_TD&rho_vs_B}
\end{figure}

Extraction of the $T_D(B_\parallel)$ dependence is complicated by the fact that the MO amplitude is affected by $g^*$, and $\gamma$ which are both  field-dependent.
Figure~\ref{fig-n2.38_TD&rho_vs_B} shows $T_D$ measured at three different $B_\perp$ fields corresponding to the beating antinode and two adjacent nodes. Despite the experimental uncertainty which becomes very large at zero field, one can conclude that to the first approximation  $T_D$  is  $B_\parallel$-independent. This behavior of $T_D$ does not resemble the monotonic  $\rho(B_\parallel)$  dependence shown in the same plot. In the framework of the interaction correction theory,  the monotonic MR is related to the reduction of the effective  number of triplets with $B_\parallel$, rather than the $1/\tau$ renormalization. The dissimilarity between the  field dependencies of $T_D$ and $\rho$   suggests that the MO damping mechanism in parallel fields is different from  that of the monotonic magnetoresistance.   This conclusion  is in line with our earlier finding \cite{alexkun_2013} where it has been demonstrated that the monotonic magnetoresistance  in the diffusive regime is not caused by the e-e interaction corrections solely.

 \subsection{$T_D^*$ dependence on the $B_\perp$ field}
According to Eq. (\ref{reduced LK+gornyi}),  the correction $\delta T_D^*$
should grow proportionally to the inverse perpendicular field. The data in Figs.~\ref{fig-TD_dr_n2.75_B0_vs_T}, and \ref{fig-TD_dr_n2.23_B3.36_vs_T} do show a certain growth of $T_D$  with $1/B_\perp$. However, once the MO amplitude is affected by the developing QHF,
the spin gap gets increased,  the amplitude of the respective oscillations is enhanced and  the corresponding
$T_D$ values are underestimated; the latter effect becomes more pronounced at stronger $B_\perp$. This results in a ``wrong''  sequence of curves in Figs.~\ref{fig-TD_dr_n10_B0_vs_T}, \ref{fig-TD_dr_n2.7_B1.98_vs_T}, and \ref{fig-TD_dr_n1.77_B3.37_vs_T} at low temperatures. In spite of the smallness of oscillations, $\delta\rho/\rho =10^{-2} \div 10^{-4}$, the interlevel interaction remain sufficiently strong to impede comparison with theory. With increasing the temperature, these effects become negligible and the regular order of the $T_D(T, B_\perp)$ curves is restored  (see the  crossing of curves in Fig.~\ref{fig-TD_dr_n1.77_B3.37_vs_T}).

 The slope  $dT_D^*/dT$  should also grow  linearly with $1/B_\perp$ (see Eq.~(\ref{reduced LK+gornyi}));
 the low density data (Fig.~\ref{fig-TD_dr_n2.75_B0_vs_T}) indeed  reflect the  growth, whereas at high densities (Fig.~\ref{fig-TD_dr_n10_B0_vs_T}) the slope is either independent of field or even has the opposite tendency. We believe  that the predicted divergency of the slope $\propto 1/\omega_c\tau$  should be cut-off in low fields by a characteristic  energy that depends on the density and parallel field.
  In general, $dT_D/dT$  and $dT_D/dB_\perp$ cannot be fitted simultaneously: if one attempts to fit the field dependence of the slope, than the calculated $T_D^*(T)$-dependence would be too steep.

\subsection{7.3. Unexpected results}
Our  study also  revealed  two effects that go beyond the existing  theory, namely, the asymmetry (skewness) of the two spin subbands, and an extra decay of the MO
amplitude with field. The major features of the two  effects are as follows:
\begin{enumerate}
\item
The factor $m^*T_D$  is noticeably different for two spin-subbands .
We assumed that $m^*$ is the same in both subbands and quantitatively characterized  the difference with a skew factor $\gamma=\left(T_{D\downarrow}-T_{D\uparrow}\right)/2$.
\item
As the temperature increases,  the skew factor  diminishes, changes sign and saturates at a small negative value.
This value becomes progressively more negative as $B_\parallel$ increases and the density decreases.
\item
In the absence of $B_\parallel$   the skew  factor grows with $B_\perp$ from zero to $\sim 10$\%  (even though the MO amplitude and  polarization factor remain small), indicating a {\em smaller} $T_D$ value in the spin majority subband.
\item
The extra decay of the MO amplitude $T_D(B_\perp)$ with $1/B_\perp$ is  irrelevant to the
interaction correction Eq.~(\ref{reduced LK+gornyi}), being either much larger (for low temperatures) or even of the opposite sign (at elevated temperatures). We modeled the extra decay with an empirical  $T_D(B_\perp)$ dependence.
\end{enumerate}

The extra decay and the skew factor are somewhat correlated.
Were we  able to trace the skew factor continuously versus $B_\perp$ field, we would have disentangled  the two parameters.
Instead, we could determine $\gamma$ only at several values of  $B_\perp$ corresponding to the nodes of beatings.

{\em  In weak fields}, the MO data can be fitted using the standard procedure with the exponential damping factor, Eqs.~(\ref{reduced LK+gornyi}), (\ref{LK-damping}), which assumes that the field-independent $\gamma=0$,  and the extra decay of the MO amplitude is negligible.
This approach, however, would leave unexplained the data at higher $B_\perp$ fields and in the vicinity of nodes.  Moreover, within such an approach, the oscillations at  lowest field overlap with the tail of the weak localization magnetoresistance, and  can be observed only over a narrow range of temperatures; both issues impede  analysis of experimental data.

Although we attributed  the step-like variation of $\gamma$ with $B_\perp$ field to the interlevel exchange interaction, we believe that
this mechanism solely cannot explain the variation of $\gamma$ with the in-plane field and temperature, and hence, there are at least two effects at work that should be considered.

\subsection{7.4. Empirical model}
Our experimental data suggest that  the existing theory of magnetooscillations in  2D interacting systems is incomplete.
We attempt therefore to sketch an empirical model that might explain the data.
The first of the aforementioned  features -- nonequivalence of the two subbands (i.e., the  large skewness $\gamma$) --
provides an evidence for the existence of a spin-direction-dependent  scattering of mobile electrons in 2D systems, which may be attributed to
a triplet state of the scatterers. Both  $\gamma$ and $d_1$ are sensitive to the relatively weak fields ($\mu_B B \ll E_F$) which indicates that the above triplet  scatterers are located in energy close to the Fermi level but do not belong to the Fermi liquid. Thus, to explain the skewness, we shall consider the picture of a two phase system consisting of the triplet localized states, and the  2D Fermi liquid coexisting and interacting with the localized states.

The  existence of the surface localized states is not surprising, they have been observed in  (a)  earlier measurements of the energy relaxation rate dominated by the
piezo-coupling of electrons with phonons at the Si-SiO$_2$ interface  \cite{reznikov_PRL_2002}, and (b) the thermodynamic measurements of the electron magnetization
\cite{reznikov_PRL_2012} which revealed the existence of collective droplets with a large spin (i.e., triplet localized states). Their existence also follows from numerous compressibility measurements which show the compressibility increase with lowering carrier density, the effect that was explained in terms of the  development of the two-phase state
upon lowering density \cite{proskuryakov_2002, fogler_2004}. The surface localized states have been considered in the earlier theories \cite{theory-localized states}.

The collective triplet localized states (we shall refer to them as large-size scatterers)
can be easily polarized in an external magnetic field \cite{reznikov_PRL_2012}. Therefore, in the absence of the parallel field  and at sufficiently low temperatures,  weak $B_\perp$ field spin-polarizes the scatterers  in the same direction as the mobile 2D electrons in the majority spin subband. Due to the Pauli principle, the parallel spins interact weaker, which should lead to a weaker scattering of carriers in the  majority  subband by the localized states, and a positive  skew factor that grows with $B_\perp$, in line with  the aforementioned observations (1) and (3).

  In this scenario
 the triplet state polarization is reduced at $T> 2\mu_B B_\perp/k_B$,
 or, alternatively, as the field decreases at a given temperature. Hence, the skew factor  must  vanish at
 $B_\perp/T \lesssim 1$\,T/K. Such anticipated $\gamma(T)$  and $\gamma(B_\perp)$ dependencies   are also in a qualitative agreement  with observations (2) and (3) (see also Figs.~\ref{fig-gamma_vs_T} and \ref{fig-6.18_Bpar}). In particular, the sharp, almost step-like changes in $\gamma (B_\perp)$ shown in Fig.~\ref{fig-6.18_Bpar} are reminiscent of the Brillouin function describing the spin magnetization of free spins \cite{reznikov_PRL_2012} at finite temperature.

Within this model, it is easy to understand why the minima of $T_D$ occur at such a high  temperature, a factor of 2 -- 3 greater than those in $\rho(T)$ \cite{klimov_prb_2008}. Indeed, the collective  triplet scatterers, due to their large size, $ > \lambda_F$,  and weak scattering potential,  produce predominantly small angle scattering. This scattering directly affects $T_D$, but contributes very little to $\rho$ determined by the large angle scattering.
The interplay between the  quantum corrections Eq.~(\ref{gornyi-damping})  (with positive $dT_D/dT$) and all-angle scattering (for the small angle scattering,  the number of scatterers grows as $1/T$ \cite{reznikov_PRL_2012})  results in the $T_D(T)$ minimum. The minimum is shifted towards higher temperatures  because the all-angle scattering is always stronger than the large-angle one. As the density decreases, the number of the large-size scatterers grows \cite{reznikov_PRL_2012} and the minima are also shifted to higher temperatures (cf. Figs. 16 and 17).

With application of the in-plane field, the surface scattering becomes weaker (and, as a consequence, the skewness vanishes); simultaneously, the $T_D(T)$ minimum  shifts to lower temperatures (see Figs. 19 and 20).
In the presence of $B_\parallel$ field, several other effects may occur. Firstly, the electrons may be redistributed between the mobile and localized subband. In this case, the frequency of  SdH oscillations would have been  $B_\parallel$-dependent. Indeed, sharp variation of the SdH density  has been reported earlier \cite{pud_granada_2004}. Secondly,  when $B_\parallel$ becomes stronger than  $B_\perp$,  the spins of both scatterers and mobile electrons should be aligned within the 2DEG plane.  However,
 because of a non-zero spin-orbit coupling, the spins of mobile electrons tend to align perpendicular to the $\textbf{k}$
 vector in the momentum space. As a result,  the difference in scattering rate between the spin-majority and -minority  subbands, $\gamma$, is expected to diminish in a strong $B_\parallel$ field. This expectation is also in a qualitative agreement  with observation (4)  (see also Fig.~\ref{gamma_vs_B_parallel}).

In the suggested model, we assume that (a)  the localized triplet states exist close to the Fermi energy, (b) these states act as potential scatterers, and (c) the SO coupling is sufficiently strong at the Si-SiO$_2$ interface. The first two assumptions are in agreement with  other available data \cite{reznikov_PRL_2012, elsewhere}. The cooldown- and sample-dependence of $\gamma$ and $d_1$ also points to their interface origin (i.e., association with shallow interface traps).
The existence of the SO coupling in Si-MOS is usually neglected
because of the large band gap and bulk inversion symmetry of Si. However, the potential well at the Si-SiO$_2$ interface is strongly asymmetric; from the magnetoresistance studies in a parallel magnetic field, we were able to estimate the strength of the SO coupling in Si-MOSFETs \cite{anisopud_prl_2002},  with a moderate value of the SO coupling parameter. Also, as we have mentioned above, our measurements  of the energy relaxation rate of 2D electrons in Si-MOSFETs \cite{reznikov_PRL_2002}  revealed a relatively strong electron-phonon coupling, which implies the piezoelectric coupling rather than the deformation potential one. The  piezoelectric coupling requires lack of the inversion symmetry at the interface, and is the prerequisite for the SO coupling.

The last (fourth)  unexpected observation is the extra decay of the oscillations, which  is important for the proper analysis of the MO amplitude.

(i) In the conventional approach, only the temperature dependence of the oscillation envelope is analyzed; this leads to  averaging  the extra decay out. This approach has been used
in the majority of previous studies (e.g., Refs.~\cite{shashkin_prl_2003, okamoto_1999}). Then, unavoidably,  the  effective mass extracted from the slope of the Dingle plot becomes dependent on the ranges of $B_\perp$  and  temperature, on the particular sample, and on the cooldown conditions through the uncontrollable density of interface traps. This  leads  to substantial scattering of the  $m^*(n)$ values measured in different experiments; this scattering exceeds by far the measurement accuracy in a single measurement run, a problem mentioned in several  experimental papers.

(ii) In this paper  we  analyzed the MO amplitude using  {\em field dependent} $T_D$.  This approach enabled us to fit reasonably well the oscillation lineshape, phase (e.g., the  locations of the minima in  $B_\perp$ field)  and the MO amplitude. As we explained above, this approach leads to the field dependent $T_D$.

(iii) Finally, as we have shown experimentally,
in the presence of a large $B_\parallel$ field, the extra amplitude damping and the spin subband skewness  vanishes, or, at least, tends to become $B_\perp$-field independent. As a result, the $T_D$ value
 becomes adequately  described with the theory Eqs.~(\ref{LK-damping}) and (\ref{reduced LK+gornyi}). For this reason, the analysis of the oscillation amplitude  in the presence of $B_\parallel$ field provides more reproducible values of $m^*(n)$. Thus the current analysis justifies our earlier conjecture   in Ref.~\cite{pud_gm_2002}, where we  measured the effective mass in the presence of the $B_\parallel$ field.

The suggested  empirical  model  provides a qualitative description of all major observations. Still, a thorough microscopic model  is required to explain  the observed discrepancy  between the measured oscillation  decay (and its sign change with temperature) and the decay calculated from Eq.~(6). The measured  skewness $\gamma$ not only decreases with $B_\parallel$, but also changes sign; this suggests that the spin-minority subband
 becomes less ``disordered'', an effect that seems puzzling.
We speculate that within the considered scenario of
easily spin-polarized triplet interface scatterers,
a smaller broadening of  levels in the spin-minority subband
may be due to a more complex multilevel structure of the energy band of collective localized states.
A more detailed theory should incorporate on  equal footing also the interlevel interaction effects (i.e., quantum Hall ferromagnetism), which as we showed are relevant even to a small MO amplitude.

Finally, we note that the existing interaction  theory describes the magneto-oscillations  of thermodynamic rather than kinetic quantities, whereas the proportionality between the  oscillations in magnetotransport and density of states, Eq.~(\ref{SdHbyLK}),  has been established only for non-interacting systems \cite{SdH}.

\section{8. Conclusions}
To summarize, we performed studies   of the Shubnikov-de Haas magnetooscillations (MO) in the interacting 2D carrier system
   in high mobility Si-MOSFETs subjected to superimposed in-plane and perpendicular magnetic fields.
We analyzed the  oscillation damping parameter $m^*T_D$ and the lineshape of oscillations
for the spin-up and spin-down subbands  as a function of temperature and both field components.

Firstly, we found that the damping parameter $m^*T_D$, to the first approximation, is the same for both spin subbands,
even though their population may differ as much as 66\%.
This implies approximate equality between  $m^*_{\uparrow}$ and $m^*_{\downarrow}$ as well as between $T_{D\uparrow}$ and $T_{D\downarrow}$.
This result suggests that the exchange interaction
between electrons takes place over the whole electron system and over a wide range of energies $\sim E_F$
(rather than within each subband and only in the vicinity of $E_F$),
regardless of how large the Zeeman splitting is.

Secondly, by analyzing the MO amplitude,  
we have shown that the experimental data systematically deviate from the conventional theory.
We stress that the deviations cannot be detected by (conventional) plotting the  MO amplitude versus temperature.
Our data indicate that the  damping factor is different for two spin subbands, and this results in skewness of the oscillation lineshape.
In the absence of the in-plane field, the damping  factor $m^*T_D$
is systematically  {\em smaller in the spin-majority subband}. The difference,
quantified by the skew factor $\gamma= (T_{D\downarrow}-T_{D\uparrow})/2T_{D0}$,  can be as large as 20\% and does not correlate with  the spin polarization degree. The skew factor
tends to decrease as $B_\parallel$ or temperature grow, or $B_\perp$ decreases. For low electron densities and high in-plane field,  $\gamma$  even changes sign. To explain qualitatively these results, we suggested an empirical  model that assumes that there is a considerable density of the easily magnetized triplet  scatterers on the $S/SiO_2$ interface.

Our results also explain the origin of the well-known problem of
strong scattering of the effective mass data. By fitting the MO amplitude  with the conventional LK formula in the low-temperature range (where $dT_D/dT$ is negative),
one obtains an underestimated effective mass, and, vice versa, the same analysis in the high-temperature range (where $dT_D/dT$ is positive) provides an overestimated $m^*$ value.
Our study shows how to avoid this ambiguity by performing the MO measurements in tilted fields and at elevated temperatures.

We compared the experimentally extracted temperature and perpendicular  field dependence of the MO damping factor   with the theory for an interacting 2D system.
The comparison revealed some qualitative similarities as well as quantitative and qualitative differences.
In accord with the theory, the  extracted $T_D$ typically grows with the temperature, with the exception of the lowest temperatures. This growth, however, is  much  weaker than the calculated dependence.
The  $T_D^*(B_\perp)$ was predicted to grow with the inverse magnetic field.
Experimentally, at low densities $T_D$ indeed increases with $1/B_\perp$ and further saturates. For high densities, $T_D$ is independent of the $B_\perp$ field, at odds with the theory.

Several  of our observations are still to be explained by a more detailed theory.
Particularly, it remains puzzling why the difference of the damping parameters in two spin subbands changes sign in the limit of large in-plane fields.
Better understanding is required to explain an interesting observation that  the  magnetic field dependence of  SdH oscillations, being at odds with
 the MO  theory in weak $B_\parallel$,   agrees surprisingly well with the same theory in stronger fields.

\section{9. Acknowledgements}
Authors are grateful to  I. S. Burmistrov, and I. Gornyi
for discussions.  M.G. and H.K acknowledge  support by the NSF grant DMR 0077825,
V.M.P. acknowledges  support by the grant 12-02-00579 from RFBR  (measurements in crossed fields),
and by 14-12-00879 from the Russian Science Foundation (SdH measurements at higher temperatures and data analysis).
 Authors also acknowledge the Shared Facility Center at LPI for using their equipment.


\begin{thebibliography}{99}
\bibitem{note-interaction} The
electron-electron interaction is commonly characterized by
 the  ratio of the Coulomb interaction energy to the Fermi
energy $E_{ee}/2E_F$, the latter is doubled due to the valley degeneracy in (100)Si. $r_s=2.63\times (10^{12}{\rm
cm}^{-2}/n)^{1/2}$ \cite{ando_review} presuming that interaction
is characterized in terms of the non-renormalized conventional band mass $m_b^0=0.19m_e$.
\end{thebibliography}
\end{document}